\documentclass[submission, Phys]{SciPost}

\usepackage{float}
\usepackage{comment}

\newcommand{\be}{\begin{equation} }
\newcommand{\ee}{\end{equation} }
\newcommand{\ba}{\begin{eqnarray} }
\newcommand{\ea}{\end{eqnarray} }

\newcommand{\bpm}{\begin{pmatrix}}
\newcommand{\epm}{\end{pmatrix}}
\newcommand{\bmm}{\begin{matrix}}
\newcommand{\emm}{\end{matrix}}

\newcommand{\la}{\label}
\newcommand{\p}{\partial}

\newcommand{\bea}{\begin{eqnarray}}
\newcommand{\eea}{\end{eqnarray}}

\begin{document}

\begin{center}{\Large \textbf{
Modulation instability in dispersive parity-broken systems
}}\end{center}

\begin{center}
Sudheesh Srivastava\textsuperscript{1,2}, Gustavo M. Monteiro \textsuperscript{3}, Sriram Ganeshan\textsuperscript{1,2, *}
\end{center}

\begin{center}
{\bf 1} Department of Physics, City College, City University of New York, New York, NY 10031, USA \\
{\bf 2} CUNY Graduate Center, New York, NY 10031\\
{\bf 3} Department of Physics and Astronomy, College of Staten Island, CUNY, Staten Island, NY 10314, USA\\
*sganeshan@ccny.cuny.edu
\end{center}

\begin{center}
\today
\end{center}


\section*{Abstract}
{\bf
This work explores the interplay between dispersive parity breaking and non-linearity in two contrasting continuous dynamical systems that exhibit Modulation Instability (MI). We begin by examining deep water odd surface gravity waves and derive the non-linear Schrödinger equation (NLSE) for the modulated envelope dynamics using the method of multiple scales. The parity breaking in the odd gravity waves results in distinct NLSEs for the right and the left mover, leading to chirality-dependent stability properties for the envelope dynamics. Moreover, the resonant interaction of gravity waves and odd viscosity-induced capillary dynamics creates a window of wave numbers in one of the chiral sectors where the envelope propagation remains stable. Following the odd gravity results, we design a one-dimensional non-reciprocal PT-symmetric dielectric model that exhibits parity-breaking effects analogous to the odd viscosity term in 2D hydrodynamics. With cubic non-linearity in the polarization dynamics, we derive the corresponding NLSE. Once again, parity breaking stabilizes the modulated envelope dynamics in the lower polariton bands. We then compare the similarities and differences between this case and that of odd gravity waves.}

\vspace{10pt}
\noindent\rule{\textwidth}{1pt}
\tableofcontents\thispagestyle{fancy}
\noindent\rule{\textwidth}{1pt}
\vspace{10pt}

\section{Introduction}
\label{sec:intro}

Parity breaking is a ubiquitous phenomenon that occurs at a microscopic scale in both synthetic and natural entities with handedness. How these microscale chiral effects translate to the collective dynamics at the macro and mesoscale systems has been an active area of research across several disciplines \cite{martin1972unified, fruchart2022oddgas, markovich2020odd, fruchart2023odd}.

One approach to studying parity-breaking effects at a macroscopic scale is to write down classical field theories such as hydrodynamics with symmetry-allowed terms in the constitutive equations encoding the chiral effects~\cite{landau1987lifshitz, fruchart2023odd}. The macroscopic approach based on symmetry has allowed the identification of equivalent physical effects across different systems. For example, odd viscosity~\cite{avron1998odd,ganeshan2017odd} is a unique term in two-dimensional fluids that preserves isotropy while breaking parity without causing dissipation. In certain broken parity two-dimensional quantum fluids, it has been investigated extensively as Hall viscosity~\cite{avron1995viscosity,tokatly2006magnetoelasticity,tokatly2007new,haldane2011geometrical,hoyos2012hall,bradlyn2012kubo,abanov2013effective, hoyos2014hall,laskin2015collective,can2015geometry,klevtsov2015geometric,scaffidi2017hydrodynamic,pellegrino2017nonlocal,berdyugin2019measuring,Monteiro2018nonresistivite,korving1966transverse}. Subsequently, this concept has been extended to various fields such as elasticity theory~\cite{scheibner2020odd}, acoustics~\cite{nassar2020nonreciprocity}, and active matter\cite{banerjee2017odd, soni2018free, markovich2020odd, monteiro2021hamiltonian, reynolds2021oddlaw}, with each system exhibiting distinct physics related to this term. In elasticity, for instance, energy must be locally supplied to achieve such parity-odd responses. The diverse class of systems where odd viscosity manifests have been extensively reviewed in Ref.~\cite{fruchart2023odd} and the references therein.  

Despite these advances, the cross-fertilization of ideas has mostly been confined to the linear-response regime, with some works investigating weakly non-linear effects in incompressible fluid systems with odd viscosity ~\cite{abanov2018odd, kirkinis2019odd, monteiro202shallow, doak2023nonlinear}. However, certain non-linear effects are also ubiquitous across various platforms and disciplines. One such phenomenon is modulation instability (MI), a fundamental concept in nonlinear wave theory, with applications across many branches of physics and engineering. MI arises due to the interplay between wave dispersion and nonlinear interactions leading to soliton formation, sideband instabilities, and in some cases Fermi-Pasta-Ulam recurrences~\cite{zakharov2009modulation}.

Historically, modulation instability (MI) was discovered almost simultaneously in surface gravity waves and optics, and it has since become a widely studied nonlinear effect~\cite{zakharov2009modulation}. In fluid dynamics, it is known as the Benjamin-Feir instability, named after its discoverers \cite{benjamin1967instability,benjamin1967disintegration}. Benjamin and Feir demonstrated that gravity waves on deep water bodies are unstable to harmonic sideband perturbations, leading to the breakdown of wave trains due to nonlinear interactions leading to the formation of more stable solitons. In optics, MI was experimentally observed as a spatial modulation instability of high-power lasers in organic solvents by Pilipetskii and Rustamov \cite{pilipetskii1965observation}, and Bespalov and Talanov published the corresponding theory \cite{bespalov1966filamentary}. 

In this work, we consider the phenomenon of modulation instability (MI) in the presence of a specific dispersive parity-breaking term for fluid dynamics and electromagnetic waves propagating in a non-reciprocal medium. For fluid dynamics, our starting point is a nonlinear irrotational odd surface wave model in the presence of gravity (a.k.a odd gravity waves) developed by two of us \cite{abanov2018odd, abanov2020variation}. We then systematically derive the Non-Linear Schrödinger Equation (NLSE) using the method of multiple scales. The parity-breaking effects result in completely different stability criteria for the left and right movers.

We also observe a resonant interaction between the gravity waves and odd surface waves in one of the chiral sectors, leading to a narrow stability window. Such resonant interaction-induced stability windows have been understood for gravity-surface tension interactions, but they symmetrically manifest in both sectors \cite{djordjevic1977two, mcgoldrick1965resonant}. This chirality-dependent stability is the hallmark of parity breaking.  The consequence of chirality-dependent instability leads to the formation of envelope solitons that propagate in only one direction starting from an arbitrary initial condition. A similar NLSE has been derived starting from a three-dimensional fluid with vertical odd viscosity \cite{doak2023nonlinear}; however, this setup is investigating three-dimensional odd viscosity and is different from our two-dimensional starting point.

 Following our analysis of odd gravity waves, a natural question arises whether light propagating in a nonreciprocal dielectric medium can exhibit an odd-viscosity-like dispersive parity-breaking phenomenon. Specifically, we aim to explore if mechanisms of parity-breaking and the resulting chiral dispersion observed in fluid dynamics can be mirrored in the electromagnetic domain, potentially revealing new insights and applications in optical materials and waveguides. To this end, we first introduce a minimal model for breaking parity symmetry in a 1D dielectric. The term that breaks parity couples the velocity of one of the oscillators to the position of the next oscillator. This term in the continuum limit contains a spatial and a time derivative which is odd under parity (P) and time reversal (T) but invariant under the combined PT action. The continuum equations of electromagnetic waves passing through this medium contain a dispersive parity-breaking term similar to the odd viscosity term arising in fluids. 
 
We show that the dispersion of the propagating waves becomes chiral in the presence of this term. Next, we investigate the effect of Kerr nonlinearity on the polarization dynamics in the context of modulation instability (MI) in such nonreciprocal systems. To analyze the MI phenomena, we derive a parity-broken nonlinear Schr\"odinger equation (NLSE) using the method of multiple scales.

There are key differences between the odd gravity wave and the non-reciprocal dielectric system. Firstly, the former is a 2D system with a 1D free surface, while the latter is a purely 1D system. In the odd gravity case, the two scalar fields are the dynamical surface profile and the scalar potential, involving only first-order temporal derivatives. In the optical case, the two fields are polarization and electric field, with dynamics governed by the second-order coupled matter Maxwell's equations in the continuum limit. The nonlinearity in this case arises from the Kerr-type cubic nonlinearity in the polarization equations.

Despite the fundamentally different underlying dynamics compared to the odd gravity fluid, the form of the NLSE is retained, and the parity-breaking term leads to an asymmetric stability criterion for left-moving and right-moving chiral sectors. Unlike the odd gravity case, the stability criterion here is entirely determined by the form of the dispersion relation, rather than the explicit form of the interaction coefficient.

This paper is organized as follows. We begin with the review of the potential theory for the odd gravity waves in Sec.~\ref{sec:hydro} and derive the odd viscosity-modified free surface dynamics. We then introduce the method of multiple scales applied to this system and derive the NLSE for the modulated envelope dynamics in Sec.~\ref{sec:multiplescales}. In Sec.~\ref{sec:stability}, we investigate the stability of the NLSE. Following the insights from the odd gravity wave phenomena, in Sec.~\ref{sec:optics}, we then develop a non-reciprocal 1D dielectric model with Kerr non-linearity. We derive the NLSE for the optical system and investigate how the interplay between parity breaking and MI manifests in the optical system.


\section{Review of a non-linear potential theory for odd gravity waves}\label{sec:hydro}
We begin by reviewing our recent work on deriving non-linear potential theory for the free surface dynamics of 2D incompressible fluids with odd viscosity and gravity\cite{abanov2018odd,abanov2020variation}. The hydrodynamic equations for incompressible fluids with odd viscosity consist of the Euler equation, together with the incompressibility condition, that is, 
\begin{align}
\p_iv_i&=0\,. \la{eq:incomp}\\
\p_tv_i+v_j\p_j v_i&=\frac{1}{\rho}\p_jT_{ij}-\p_i(gy)\,,  \la{eq:euler}
 \end{align} 
Here,  $v_i$ are the components of the velocity field, $\rho$ is the constant and uniform fluid density, and the summation over repeated indices is assumed ($i, j=1,2$).  The term $g y$ is the external gravitational potential and  $T_{ij}$ is the the stress tensor. For a fluid with only odd viscosity, we have
\begin{align}
   T_{ij} = -p\delta_{ij} +\nu_o \rho\, (\p_i^*v_j+\p_i v_j^*)\,.
 \la{eq:Tij}  
\end{align}
The first term of the stress tensor~(\ref{eq:Tij}) is the standard isotropic pressure term. The second term in Eq.~(\ref{eq:Tij}) is the odd viscosity term. The coefficient $\nu_o$ is known as kinematic odd viscosity (or Hall viscosity). In writing this term we introduced the notation $a_i^* \equiv \epsilon_{ij}a_j$ so that the ``starred'' vector $\mathbf{a}^*$ is just a vector $\mathbf{a}$ rotated by 90 degrees clockwise. 

Under the incompressibility condition~(\ref{eq:incomp}), we have $\Delta v_i^*=\p_i\omega$ and using this identity Eq.~(\ref{eq:euler}) becomes
\be
\p_t v_i+v_j\p_j v_i=-\p_i\left(\frac{\tilde p}{\rho}+gy\right)\,. \la{eq:euler2}
\ee
This differs from the ordinary Euler equation in the definition of the modified pressure~\cite{avron1998odd, ganeshan2017odd} $\tilde p=p-\nu_o\rho\,\omega\,$ where $\omega=\p_i v_i^*$ is the fluid vorticity. 

\subsection{ Irrotational limit of the free surface dynamics and boundary layer approximation}
\begin{figure}
    \centering
    \includegraphics[scale=0.4]{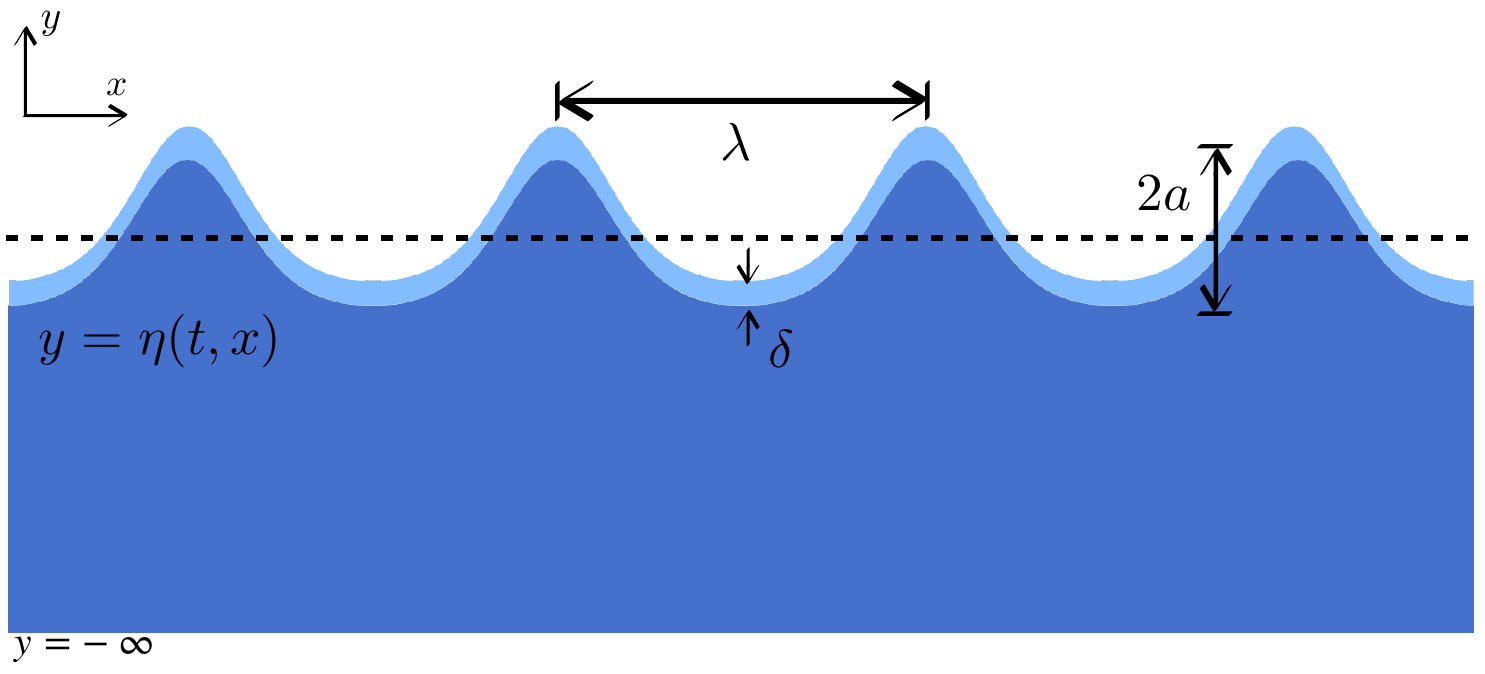}
    \caption{Schematics of the deep water system with free surface. Here, $a$ denotes the amplitude of the edge profile, $\delta$ is the boundary layer thickness and $\lambda$ is the wavelength.}
    \label{fig:schematic}
\end{figure}

We now review how one can use boundary layer theory to write effective free surface dynamics of the odd gravity waves only in terms of the scalar potential $\theta$ and the free surface profile $\eta$. We consider the fluid domain as shown in Fig.~\ref{fig:schematic}. The free surface is a dynamical interface $y=\eta(t,x)$ between two fluids (water and air) where we impose one kinematic and two dynamical boundary conditions. The kinematic boundary condition states that the velocity of the fluid normal to the boundary is equal to the rate of change of the boundary shape. 
\begin{align}
\p_t\eta=\left(v_y-v_x\p_x\eta \right)\big|_{y=\eta(t,x)}\,.
\end{align}
The pair of dynamical boundary conditions $n_iT_{ij}\big|_{y=\eta(t,x)}=0$ impose that there are no normal and tangent forces acting on an element of the fluid surface, where, $n_i$ are the components of the normal vector to the surface $y=\eta(t,x)$. In our previous works, we have shown that the presence of two dynamical boundary conditions, together with incompressibility, requires a singular boundary layer where the divergent vorticity is confined \cite{abanov2018odd, abanov2019hydro}. The role of this singular boundary layer is to ensure that there are no tangential forces on this interface.

We have shown that regardless of the boundary layer mechanism — dissipative \cite{abanov2018odd} or compressible \cite{abanov2019hydro} — the normal component of the dynamical boundary condition (DBC) is universal and geometric \cite{abanov2018free}. Assuming that the boundary layer is stable and is confined to short length scales, the free surface problem can be written as an effective description of the fluid with the effects of the boundary layer encoded in the odd viscosity modified pressure term,
\begin{align}
	\tilde p\big|_{y=\eta(t,x)}=(p-\nu_o\rho\,\omega)\big|_{y=\eta(t,x)}= \frac{2\nu_o\rho}{\sqrt{1+(\p_x\eta)^2}} \p_xv_n \big|_{y=\eta(t,x)}\,. \la{DBC}
\end{align} 
Here $v_n$ is the velocity component which is normal to the boundary, taken at $y=\eta(t,x)$. Only after taking care of the tangent stress boundary condition and taking the thin boundary layer limit, we can define velocity field as a gradient of a scalar function $v_i=\p_i\theta$. In the following, we set the density $\rho=1$, for simplicity. 

 The bulk equation for the irrotational system is simply the Laplace equation for the potential 
 \begin{align}
 	\Delta \theta=0,\quad -\infty<y<\eta(t,x)
 \end{align}
The boundary conditions at the top and bottom of the fluid domain can be written as,
\begin{align}
       &\theta_y=0\,, \qquad \qquad \qquad \qquad y=-\infty, \la{eq:bottomBC1}
    \\
    &\eta_t+\eta_x\theta_x=\theta_y\,, \qquad \qquad y=\eta(t,x), \la{eq:KBC1}
    \\
    &\theta_t+\frac{\theta_x^2+\theta_y^2}{2}+g\eta=\frac{2\nu_o}{\sqrt{1+\eta_x^2}}\,\p_x\left[\frac{\eta_t}{\sqrt{1+\eta_x^2}}\right],  \;\, y=\eta(t,x). \la{eq:DBC1}
\end{align}
 Here, Eq.~(\ref{eq:KBC1}) is the kinematic boundary condition and Eq.~(\ref{eq:DBC1}) is the odd viscosity modified dynamic boundary condition on the top surface, whereas Eq.~(\ref{eq:bottomBC1}) simply states that $v_y=0$ in the bottom surface. Note that the boundary condition at the bottom surface is a slip boundary condition. For the deep water problem, a more convenient set of variables is given by $(x,y,t)\rightarrow(x,\chi,t)$ where $\chi=y-\eta(x,t)$. The advantage of these variables is that it removes the implicit dependence of the $\theta$ on the free surface profile $\eta$ in the boundary conditions. On the other hand, we pay the price by making the bulk Laplace equation non-linear. In terms of these variables, the bulk equation is given by
 \begin{align}
 	\theta_{xx}+(1+\eta_x^2)\theta_{\chi\chi}-2\p_x(\eta_x\theta_\chi)+\eta_{xx}\theta_{\chi}=0, \qquad \qquad -\infty\leq \chi\leq 0
 	\label{eq:bulkchi}
 \end{align}
 \begin{figure}
    \centering
    \includegraphics[scale=0.35]{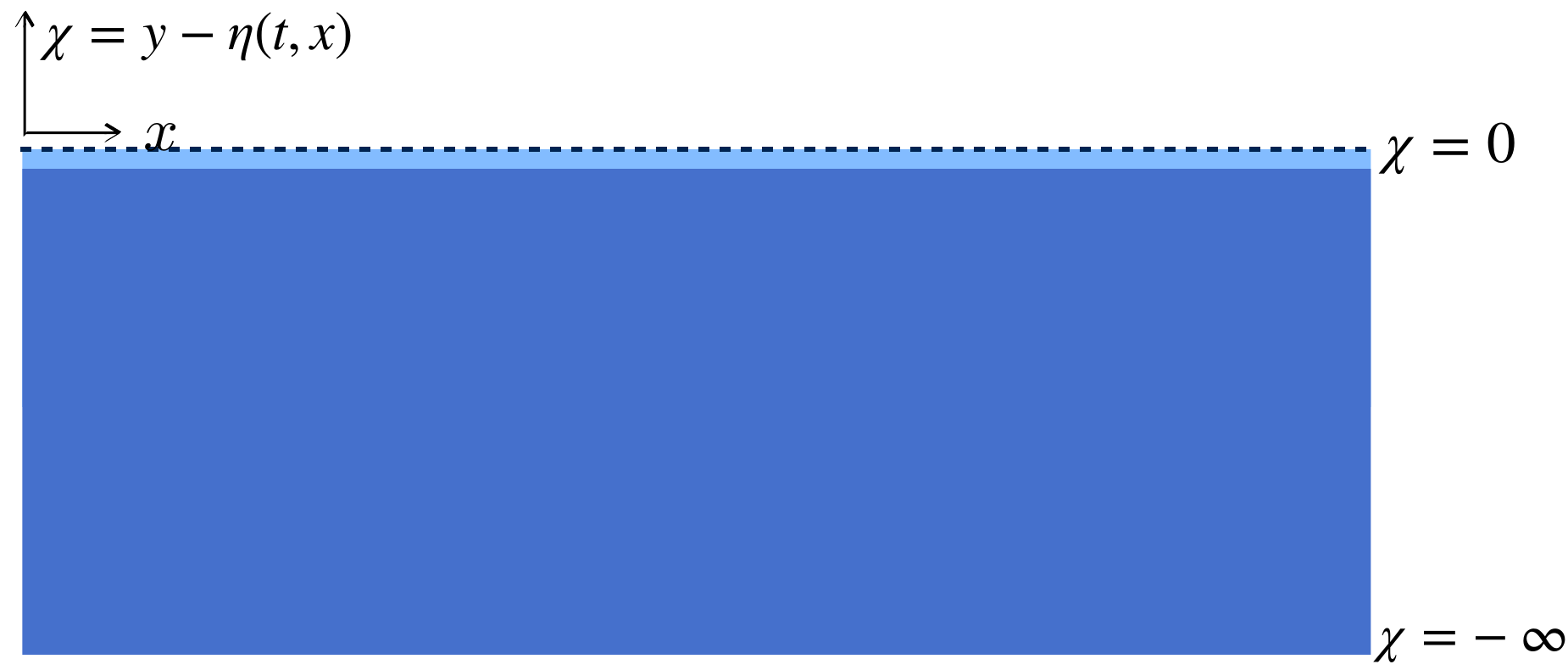}
    \caption{Schematics of the deep water system with free surface in the the new coordinate system, where the vertical coordinate $y$ is transformed to $\chi=y-\eta(t,x)$, resulting in a flat top surface.}
    \label{fig:schematicflattop}
\end{figure}
The schematic of the free surface system in the $(x,\chi,t)$ coordinates is shown in Fig.~\ref{fig:schematicflattop}. The equations at the top and bottom boundary can be written as
 \begin{align}
       &\theta_\chi=0\,, \qquad \qquad \qquad\qquad \qquad \,\,\,\,  \chi=-\infty, \label{eq:bottomBC}
    \\
    &\eta_t+\eta_x\theta_x=(1+\eta_x^2)\theta_\chi\,, \qquad \qquad \chi=0, \label{eq:KBC}
    \\
    &\theta_t-\eta_t\theta_\chi+\frac{\theta_x^2+\theta_\chi^2-2\eta_x\theta_\chi\theta_x+\eta_x^2\theta_\chi^2}{2}+g\eta=\frac{2\nu_o}{\sqrt{1+\eta_x^2}}\,\p_x\left[\frac{\eta_t}{\sqrt{1+\eta_x^2}}\right],  \;\, \chi=0. \label{eq:DBC}
\end{align}
where the subscript notation corresponds to the action of partial derivative (i.e. $\theta_{x} := \partial_{x}\theta$). In the following, we will focus on the weakly non-linear analysis using the method of multiple scales of the above system of equations.   

\section{Method of multiple scales for envelope dynamics and non-Linear Schr\"odinger equation}\label{sec:multiplescales}

The effect of modulation instability results from the interaction between a strong carrier harmonic wave and small sideband perturbations. This is a particular case of four-wave interaction, where the weak modulation imposed on a harmonic wave leads to the dynamics of the modulated envelope at much slower time scales. The slow dynamics of the envelope can be accessed by the method of multiple scales \cite{hasimoto1972nonlinear}, and due to the non-linear interactions of the waves, it can lead to the unstable growth of the sideband perturbations. These instabilities can be a precursor to the formation of more stable entities, such as solitons. To find envelope dynamics for the system, we introduce multi-scale variables
\begin{align}
	\sigma=kx-\omega t\,,\quad \xi= \epsilon (x-c_g t )\,,\quad \tau =\epsilon^2 t \label{multiscale}
\end{align}
where $k$ is the arbitrary wavenumber, $\omega$ is the dispersion, $c_g=d\omega/dk$ is the group velocity of the wave-packet and $\epsilon \ll 1$ is the wave steepness parameter. Furthermore, we seek asymptotic solutions for the system of equations~(\ref{eq:bulkchi}-\ref{eq:DBC}) that will lead to the envelope dynamics for the slow degrees of freedom, that is,
\begin{align}
	\theta(\sigma, \xi,\chi,\tau)&=\epsilon\,\theta_1(\sigma, \xi,\chi,\tau)+\epsilon^2\,\theta_2(\sigma, \xi,\chi,\tau)+\epsilon^3\,\theta_3(\sigma, \xi,\chi,\tau) +..., \\
	 \eta(\sigma, \xi,\tau) &=\epsilon\,\eta_1(\sigma,\xi,\tau)+\epsilon ^2\,\eta_2(\sigma, \xi,\tau)+\epsilon^3\,\eta_3(\sigma, \xi,\tau) +...
	 \label{eq:multiscale1}
	\end{align}
At every order, the fields are further expanded to the relevant number of harmonics that are expected to be generated due to the non-linear interactions,
\begin{align}
	\theta_{j}(\sigma,\xi,\chi,\tau)=\sum_{n=-j}^j \theta_{jn}(\xi,\chi,\tau) q^n,\quad \eta_{j}(\sigma, \xi,\tau)=\sum_{n=-j}^j \eta_{jn}(\xi,\tau)q^n, \quad q=e^{i  \sigma }.
	\label{eq:multiscale2}
\end{align}

Variables $\theta_{j}$ and $\eta_{j}$ are the perturbative expansion of fields which contributes to the solution at different orders in nonlinearity. They are further expanded in ($\theta_{jn}$ and $\eta_{jn}$) corresponding to the various possible harmonics generated at that order. This form is motivated by the fact that higher harmonics become relevant to dynamics processes at higher orders. A systematic calculation involves comparing terms at every order of $\mathcal{O}(\epsilon^n q^m)$ for both bulk and boundary equations. All the higher harmonics are written in terms of the slow envelope amplitude at the first order $\eta_{11}(\xi, \tau)$ and at the order $\mathcal{O}(\epsilon^3 q^1)$, we obtain the NLSE as a solvability condition on  $\eta_{11}(\xi, \tau)$. Below we outline the order-by-order results relegating the details of this calculation to the Appendix~\ref{app:hydro}.
\subsection{Linear order}
\begin{figure}
    \centering
    \includegraphics[width=0.5\paperwidth]{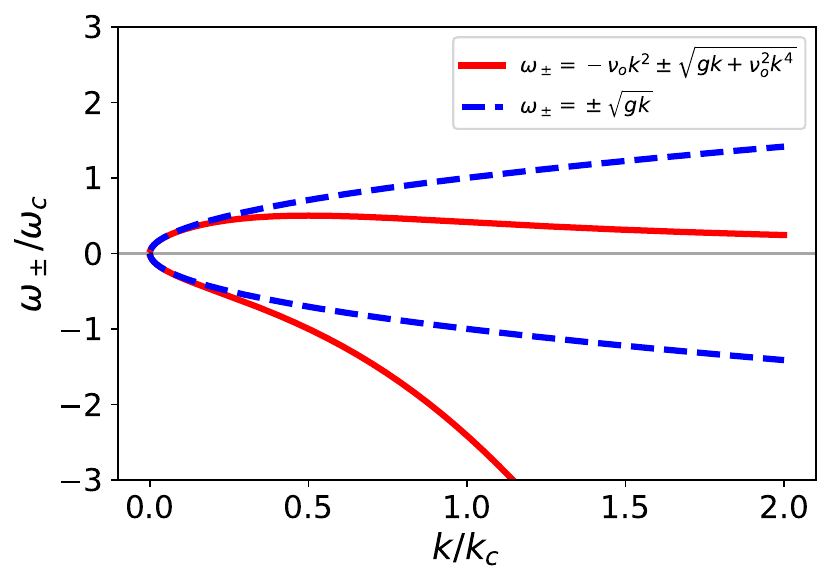}
    \caption{Dispersion for the gravity waves and odd-gravity waves $\omega_{\pm}$. The dashed curve corresponds to the gravity waves $(\nu_o=0)$. The characteristic scales $k_c$ and $\omega_c$ are defined in terms of $\nu_o$ and $g$, such that, $k_c=\left(\tfrac{g}{\nu_o^2}\right)^{1/3}$ and $\omega_c =\nu_o k_c^2$.}
    \label{fig:dispersion}
\end{figure}

Plugging the multiscale ansatz defined in Eqs.~(\ref{eq:multiscale1}) and (\ref{eq:multiscale2}) into Eqs.~(\ref{eq:bulkchi}-\ref{eq:DBC}) and comparing terms at the order $\mathcal{O}(\epsilon \,q)$, we obtain,
\begin{align}
	\p_{\chi}^2\theta_{11}&=k^2 \theta_{11},\quad -\infty\leq \chi\leq 0\\
		 -i\omega \eta_{11}&= \p_\chi \theta_{11}\,\,\,\qquad \chi=0\\
	  -i\omega \theta_{11}+g \eta_{11} &= 2\nu_o \omega k \eta_{11}, \quad \chi=0\\
	  \p_\chi \theta_{11}&=0,              \qquad      \quad    \quad \chi=-\infty
	\end{align}
	We consider $k>0$ from here on. 	For the $\mathcal{O}(\epsilon)$ order, we additionally have $\theta_{10}=A_0(\xi, \tau)$ and $\eta_{10}=0$. $A_0(\xi, \tau)$ is an arbitrary function that is fixed at higher order.  Solving these equations and boundary conditions, in the linear order we obtain:
 \begin{align}
     \theta_{11}=A(\xi,\tau)e^{k\chi} \qquad \eta_{11}=i (k/\omega) A(\xi,\tau) \label{linear_1_harmonic}
 \end{align}
 where $A(\xi,\tau)$ is the slowly varying amplitude of the envelope whose dynamics in these slow variables will be fixed to be NLSE at the $\mathcal{O}(\epsilon^3 q)$.  Solving the equation above, we obtain the dispersion relation for the odd gravity waves~\cite{abanov2018odd}, 
\begin{align}
    \omega_{\pm}=-\nu_o k^2\pm\sqrt{g k+\nu_o^2k^4}\,.
\end{align}
This dispersion has a different form for the two sectors indicating parity breaking symmetry. In the zero gravity limit ($g\rightarrow 0$) the effects of the parity breaking are maximal, where we observe that the right-movers tend towards $\omega_+\rightarrow 0$, while left-movers have a parabolic dispersion $\omega_-\rightarrow -2\nu_o k^2$ as shown in Fig.~\ref{fig:dispersion}. 
\subsection{Higher order non-linear terms and envelope dynamics}

The higher-order calculations are a bit unwieldy so we only present the final answers with the details relegated to the Appendix \ref{app:hydro}. At second order and first harmonic $\mathcal{O}(\epsilon^2 q)$ we get:
 \begin{align}
     \theta_{21}=A_2(\xi,\tau)e^{k\chi}-iA_{\xi}\chi e^{k\chi}, \qquad \eta_{21}=i (k/\omega) A_2+\left(\frac{\omega-c_g k}{\omega^2}\right) A_{\xi}\,,\label{nonlinear_1_harmonic}
 \end{align}
 where $A_2(\xi,\tau)$ is arbitrary and remains undetermined since this term always appears in a particular combination with the $\eta_{21}$. This also generates second harmonics due to non-linear interactions present at this order, which are given by
 \begin{align}
     \theta_{22}=\left(\frac{-6i\nu_o k^4}{\omega(\omega+6\nu_o k^2)}\right)A^2 e^{2k\chi}+i\left(\frac{k^2}{\omega}\right)A^2 e^{k\chi} \qquad \eta_{22}=\left(\frac{-k^3}{\omega(\omega+6\nu_o k^2)}\right) A^2.\label{nonlinear_2_harmonic}
 \end{align} 
  Note that the second-harmonic solution has a resonance in one of the sectors when the condition $\omega_-+6\nu_o k^2 =0$ is satisfied. At this value of $k$, our expansion breaks down and one needs a different counting scheme in $\epsilon$ near this value of $k$, which is beyond the scope of  this paper. This resonance also manifests in the gravity-capillary waves for a particular value of $k$ and has been addressed by McGoldrick~\cite{mcgoldrick1965resonant,mcgoldrick1970experiment}. 
 
 At the third order, that is, $\mathcal O(\epsilon^3)$, the bulk equations are solved for $\theta_{30}$ and $\theta_{31}$, which gives us
 \begin{align}
      \theta_{30}(\xi,\chi,\tau)=\left(\frac{-\p_{\xi\xi}^2A_0}{2}\right)\chi^2 + \left(\frac{k^2c_g}{\omega^2} + \frac{k^2 \chi}{\omega}\right) e^{k\chi} \p_{\xi\xi}^2 |A_1|^2.
  \end{align}
The boundary condition imposes ($     \p_{\chi}\theta_{30}=0 $ as $\chi \rightarrow -\infty$) leading to the constraint $A_0=0$ \footnote{Note that this is not the case if we use the finite depth fluid and we need to express $A_0$ in terms of $A$ and its derivatives in a consistent way}.
We further need the bulk expression for the $\theta_{31}$ fields to substitute them into the two surface equations at the order $\mathcal{O}(\epsilon^3 q^1)$. The two surface equations in this order can be viewed as a system of two linear equations for the fields $\theta_{31}(\xi,0,\tau)$ and $\eta_{31}(\xi, \tau)$.  Solving this system at the third order, we finally arrive at a Non-Linear Schr\"odinger Equation (NLSE) for the envelope amplitude $A(\xi,\tau)$. This NLSE turns out to be the consistency condition for the perturbative solution to hold and is typical of how the method of multiple scales works. The coefficients of NLSE now depend on parity-breaking parameter $\nu_o$, suggesting different stability criteria of the envelope dynamics for the right propagating NLSE and the left propagating NLSE. The general form of the NLSE can be written as
\begin{align}
    iA_{\tau}+L A_{\xi\xi}-M A|A|^2=0\,,\label{eq:nls-full}
\end{align}
where we have defined
\begin{align}
	L= \frac{1}{2} \omega''(k),\qquad M=\frac{2k^4}{\omega}\left(1-\frac{3\nu_o k^2\omega}{(\omega+6\nu_o k^2)(\omega+\nu_o k^2)}\right).
\label{eq:nls-full-LM}
\end{align}

The parity-breaking effect of odd viscosity enters in both the dispersion relation $\omega(k)$ and in the interaction coefficient $M$.  Before we discuss the stability analysis of the NLSE for the general case, let us take the limit when $\nu_o\rightarrow0$ and compare it with the NLSE obtained for the gravity waves. In this regime, the dispersion relation becomes $\omega=\omega_{\text{gravity}}=\pm\sqrt{gk}$, the interaction term becomes $M=2 k^2/\omega_{\text{gravity}}$ and the NLSE takes the form
\begin{align}
	  iA_{\tau}+\frac{1}{2}\omega''_{\text{gravity}} A_{\xi\xi}-\frac{2k^2}{\omega_{\text{gravity}}}A|A|^2=0,\qquad \nu_o\rightarrow 0.\label{eq:nls-gravity}
\end{align}
This equation is consistent with the NLSE for the deep water gravity waves \cite{hasimoto1972nonlinear,davey1974three}.  

In the opposite limit of $g\rightarrow 0$, we still have propagating odd surface waves. They fully break parity symmetry, which manifests in completely different envelope dynamics depending on whether the choice of sector, i.e., $\omega=\omega_{\pm}$.  First, we will consider the $\omega_-$ case and take the $g\rightarrow 0$ limit. In this case the dispersion relation becomes $\omega_-=-2\nu_o k^2$ and consequently $L=-2\nu_o$ and $M=k^2/(2\nu_o)$ leading to the following NLSE,
\begin{align}
	  iA_{\tau}-2\nu_o A_{\xi\xi}-\frac{k^2}{2\nu_o}A|A|^2=0\qquad g\rightarrow 0,\,\, \omega=\omega_-.\label{eq:nls-oddkneg}
\end{align}
On the other hand, taking the $g\rightarrow 0$ limit is a bit pathological for the $\omega_+$ sector since the dispersion $\omega_+=0$ becomes a zero mode. This causes the interaction term $M$ to diverge and the dispersion term $L=0$. We can consider the leading order behavior in the small $g$ limit as a regulator and we obtain
\begin{align}
	  iA_{\tau}+\frac{1}{2}\left(\frac{g}{\nu_o k^3}\right)A_{\xi\xi}-\frac{4 \nu_o k^5}{g}A|A|^2=0\qquad g\rightarrow 0,\,\, \omega=\omega_+\,.\label{eq:nls-oddkpos}
\end{align}

In fact, this forces the amplitude to vanish. However, scaling the amplitude by $\sqrt{g}$, that is, $A=\sqrt{g} \tilde A$, and taking the $g\rightarrow 0$, we find that 
\begin{align}
	  i\tilde A_{\tau}-4 \nu_o k^5|\tilde A|^2\tilde A=0\,.\label{eq:nls-oddkpos2}
\end{align}
This shows that the envelope amplitude becomes asymptotically proportional to $\sqrt{g}$, with the proportionality constant $\tilde A$ satisfying Eq.~(\ref{eq:nls-oddkpos2}). The absence of the dispersion term keeps $\tilde A$ bounded, since Eq.~(\ref{eq:nls-oddkpos2}) can be solve analytically, giving us
\begin{align}
    \tilde A=\tilde A_0\,e^{-i4\nu_ok^5|\tilde A_0|^2\tau}\,,
\end{align}
for some constant $\tilde A_0$.

\subsection{Stability of the NLSE and soliton formation}\label{sec:stability}

We now consider the linear stability of the envelope dynamics for the NLSE derived in Eqs.~(\ref{eq:nls-full}) and (\ref{eq:nls-full-LM}). The stability of the envelope dynamics is deduced by expanding the envelope around Stokes solution $ A=A_0 e^{-i M |A_0|^2 \tau}$. Let us consider infinitesimal perturbations to the solution:
\begin{equation}
    A= A_0 (1 + a) e^{i(\mu - M |A_0|^2 \tau)}
\end{equation}
 Substituting the ansatz above in the NLSE, and keeping first-order terms in $a$ and $\mu$, we get two equations, which correspond to the real and imaginary parts
\begin{align}
    \partial_{\tau} a + L \partial_{\xi}^{2} \mu=0,\qquad
      \partial_{\tau} \mu - L\partial_{\xi}^2 a + 2M |A_0|^2 a =0.
\end{align}
This linear system of equations can be expanded in modes $(a, \mu)=(\kappa_a, \kappa_{\mu}) e^{i(\delta \xi - \lambda \tau)}$ and the resulting system can be cast into a matrix form
\begin{equation}
     \begin{pmatrix}
  i\lambda &  \delta^2 L\\ 
  (2M |A_0|^2 + \delta^2 L) & -i\lambda
\end{pmatrix}  \begin{pmatrix}
  \kappa_a \\ 
  \kappa_\mu
\end{pmatrix} =0\,.
\end{equation}
This expression admits real solutions for $\lambda$ when 
\begin{equation}
    L(2 M |A_0|^2 +\delta^2 L) \geq 0\,.
    \label{NLS_STABILITY}
\end{equation}

The perturbation remains bounded when $\lambda$ is real and the Stokes' solution remains stable. On the other hand, when the condition~(\ref{NLS_STABILITY}) is not satisfied the perturbation grows in an unbounded fashion. In this case, the harmonic perturbation on the mode grows with growth rate $\gamma$, which is given by 
\begin{align}
    \gamma =|\delta|\sqrt{-2ML|A_0|^2-\delta^2 L^2}.\label{eq:growth_rate}
\end{align}
\begin{figure}
    \centering
    \includegraphics[width=0.24\paperwidth]{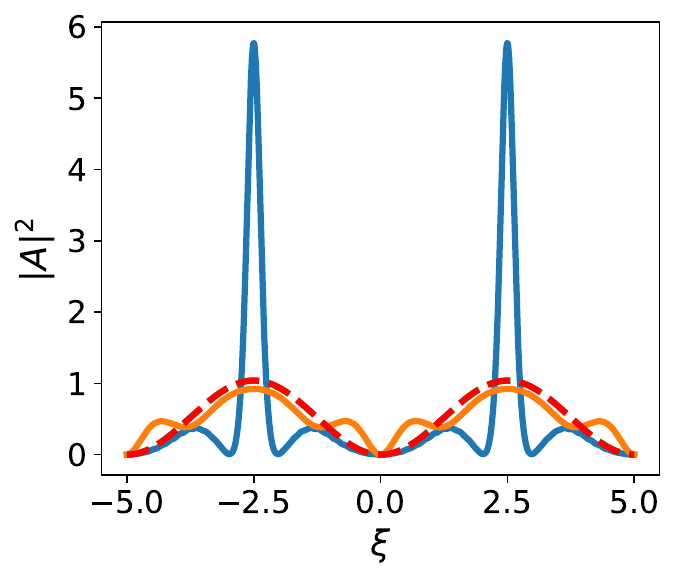}
  \includegraphics[width=0.42\paperwidth]{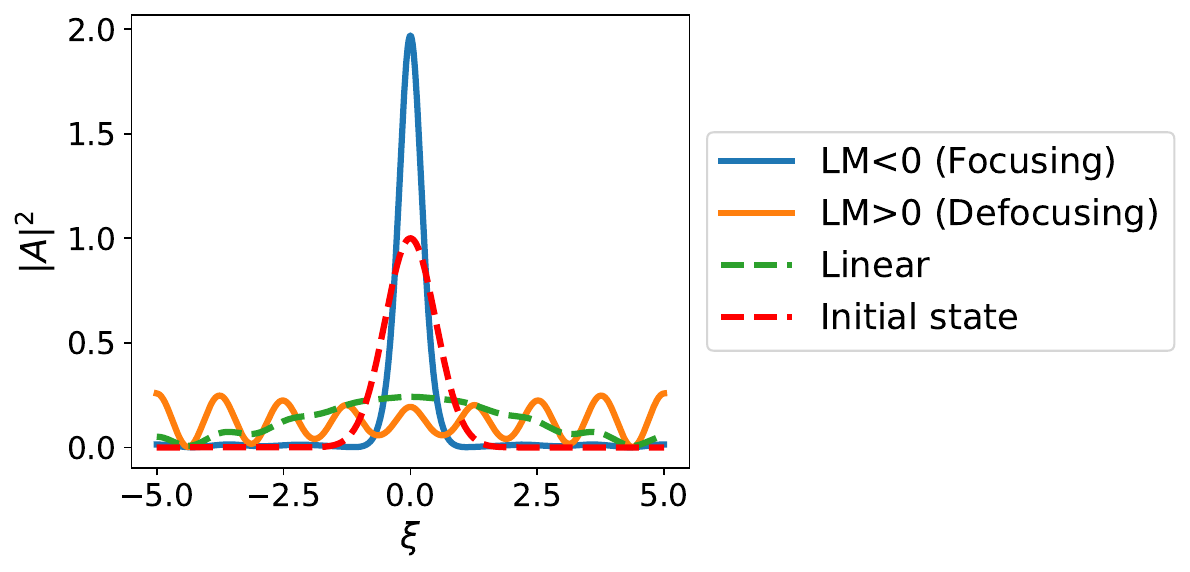}
    \caption{Time evolution of initial wave packet in two regimes of the NLSE. The final state at $\tau =2$ for the two cases has been contrasted. (Left) Initial state consisting of a primary harmonic with $\bar k= \pi/5$ and sideband perturbations in bandwidth $\delta = 0.001$, $A(\xi,0)=\sin{\bar k\xi}+0.01\left(\sin{(\bar k +\delta)\xi}+\sin{(\bar k -\delta)\xi}\right)$. (Right) Initial Gaussian wave packet, $A(\xi,0)=e^{-\xi^2}$.}
    \label{fig:soliton} 
\end{figure}
In the zero bandwidth regime of the perturbation, that is, $\delta\rightarrow 0$ limit, the criterion for the envelope stability under modulations in amplitude (or phase) simplifies to 
\begin{align}
    LM\geq 0.
\end{align}

In Fig.~\ref{fig:soliton}, we investigate the initial value problem for the NLSE with $LM > 0$ and $LM < 0$, using a perturbed plane wave and a Gaussian packet as the initial conditions.  For the case where $LM > 0$, we observe a ``superlinear" spreading of the initial profile, while for the unstable case where $LM < 0$, we notice the onset of soliton formation.

In the vanishing bandwidth limit, the stability of the NLSE only depends on the relative signs of the  $L$ and $M$ terms. We now analyze the relative signs of these for the expressions given in Eq.~(\ref{eq:nls-full-LM}). A compact way of studying the stability criterion for different parameter regimes is in terms of dimensionless variables $\bar \omega=\omega/\omega_c$, $\bar k=k/k_c$ and $\bar A=\nu_o A$, defined in terms of the characteristic scales $ k_c=\left(\frac{g}{\nu_o^2}\right)^{1/3}$, $\omega_c =\nu_o k_c^2$. In terms of the dimensionless variables, we see that the small wavenumber limit $(\bar k \ll 1)$ corresponds to the gravity-dominated regime and the large wavenumber one $(\bar k\gg 1)$ is the odd viscosity dominated regime, where parity breaking effects are most pronounced. The interpolating regime $\bar k\sim 1$ is where the gravity effects are comparable to that of the odd viscosity effects. There is also a resonance present due to the interaction between the gravity waves and odd surface waves in the intermediate regime $\bar k\sim 1$ albeit only in the $\omega_-$ sector. 

\begin{figure}[H]
    \centering
    \includegraphics[width=0.34\paperwidth]{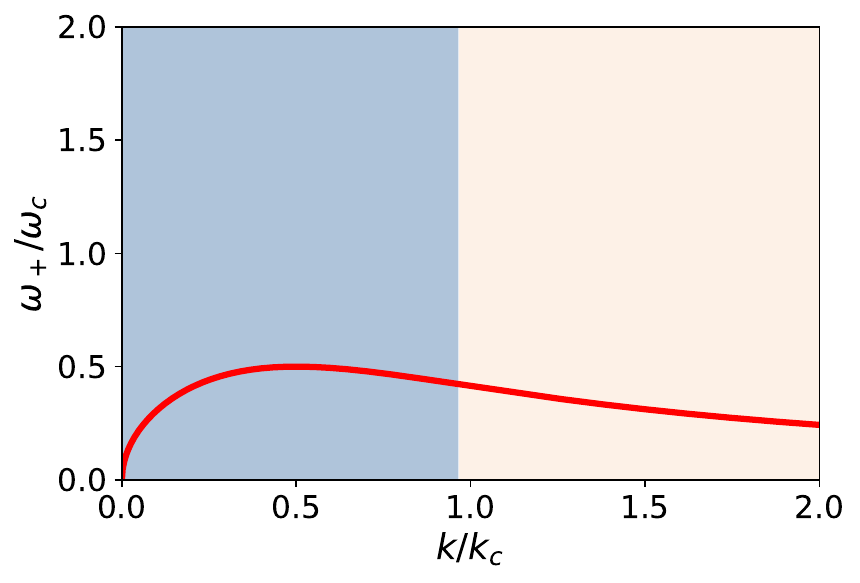}
  \includegraphics[width=0.36\paperwidth]{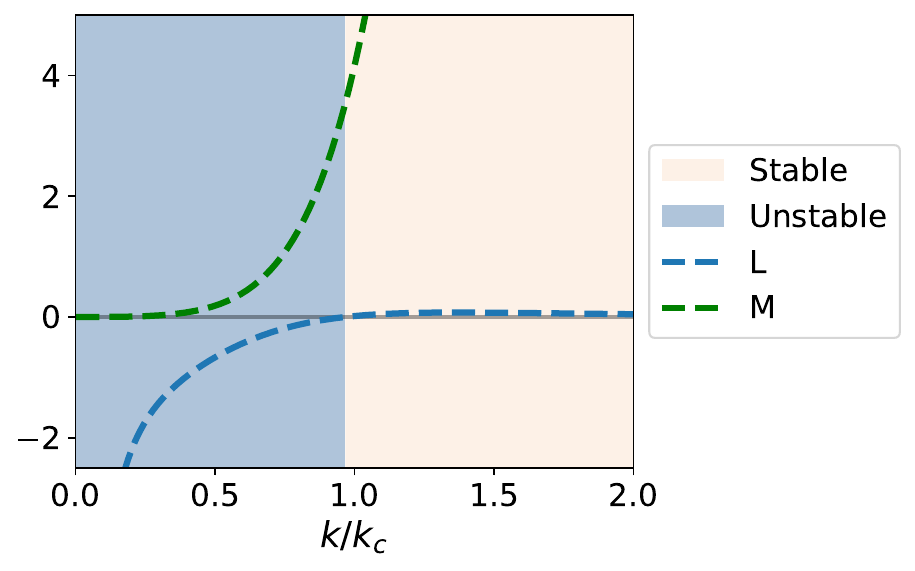}
    \caption{Left: Stability for the $\omega_+$ sector of the odd-gravity waves, calculated in vanishing bandwidth $\delta \rightarrow 0$. Right: Coefficients of NLSE.  In this sector, the band curvature changes the sign for $\bar k>0.965$ and solely determines the onset of stability since the nonlinearity doesn't change the sign.}
    \label{fig:odd_limit-plus} 
\end{figure}

\begin{figure}[H]
    \centering
    \includegraphics[width=0.34\paperwidth]{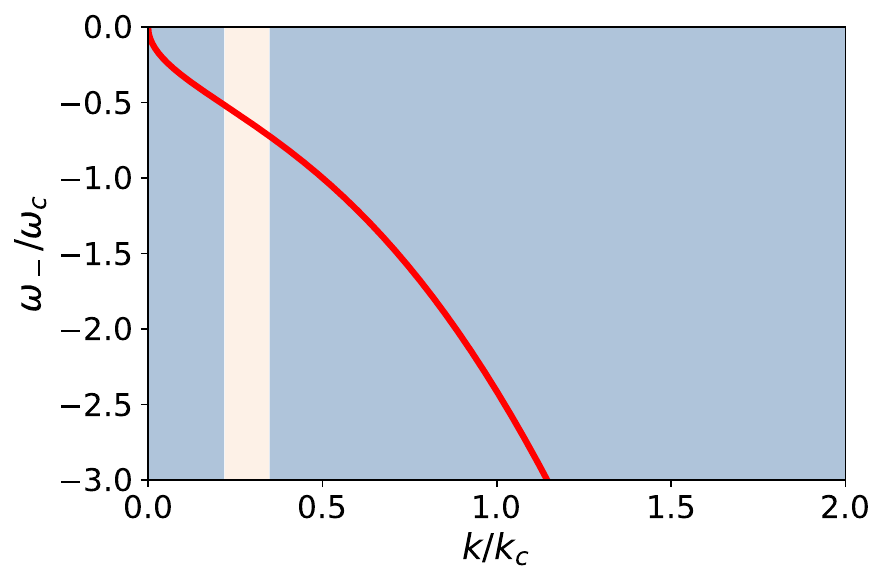}
  \includegraphics[width=0.355\paperwidth]{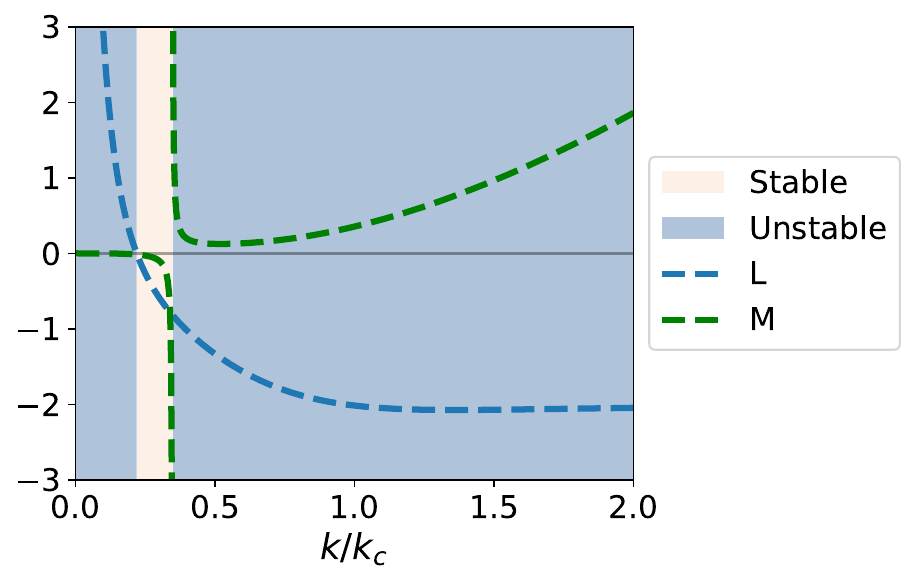}
    \caption{Left: Stability for the $\omega_-$ sector of the odd-gravity waves, calculated in vanishing bandwidth $\delta \rightarrow 0$. Right: Coefficients of NLSE. In this sector, the band curvature changes sign for $\bar k>0.218$. The nonlinearity changes sign for $\bar k> 0.346$. The window $0.218 < \bar k < 0.346$ is stable to MI. }
    \label{fig:odd_limit-minus}   
\end{figure}

It is straightforward to see that the gravity-dominated regime ($\bar k\ll 1$) with $L M=-\bar k^2/4<0$ is unstable for both $\omega_{\pm}$ sectors. On the other hand, we observe a sector-dependent stability in the odd viscosity-dominated regime of $\bar k\gg 1$. For the $\omega_-$ sector, we have $L M=-\bar k^2<0$ which is unstable and leads to soliton formation. However, the  $\omega_+$ sector in this regime is fully stable with $LM=2\bar k^2>0$ (see Figs.~\ref{fig:odd_limit-plus} and ~\ref{fig:odd_limit-minus}). The physical consequence of such a strong parity breaking is that soliton formation post instability will only manifest for the  $\omega_-$ and the $\omega_+$ may become ``superlinear" spreading of an initially localized perturbation. 

The intermediate regime of $\bar k\sim 1$ shows non-trivial interaction between the gravity waves and odd surface waves and it is in this regime where the stability criterion changes.  
For the $\omega_+$ sector, the stability arises because of the change in the sign of the band curvature at $\bar k=0.965$. We calculate this by finding the real root of the equation $\bar \omega^{''}(\bar k)=0$ for $\omega_+$. The interaction term $M$ does not change its sign in this sector as shown in Fig.~\ref{fig:odd_limit-plus}.

The $\omega_-$ sector is interesting because there is a window of stability that emerges in this sector due to the sign change in the band curvature at $\bar k=0.218$  and also a resonance-induced sign change in the interaction term $M$ at $\bar k=0.346$ (see Fig.~\ref{fig:odd_limit-minus}). As discussed before, near this resonance our original counting scheme breaks down and a different counting scheme needs to be used. Similar resonance also manifests in the gravity capillary interaction, however, it happens for both sectors. These two consecutive sign changes in $L$ and $M$ for nearby values of $\bar k$ lead to a narrow window of stability in the range ($0.218 < \bar k < 0.346$) as shown in Fig.~\ref{fig:odd_limit-minus}. 

\begin{figure}[H]
    \centering
    \includegraphics[width=0.275\paperwidth]{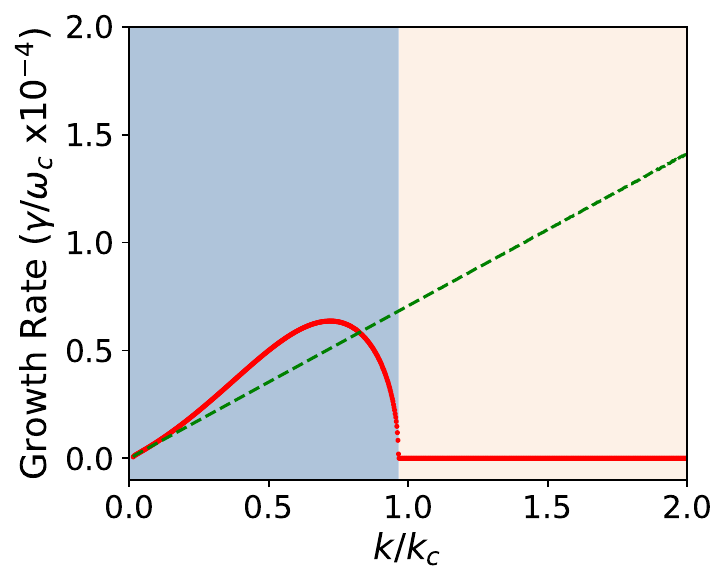}
    \includegraphics[width=0.333\paperwidth]{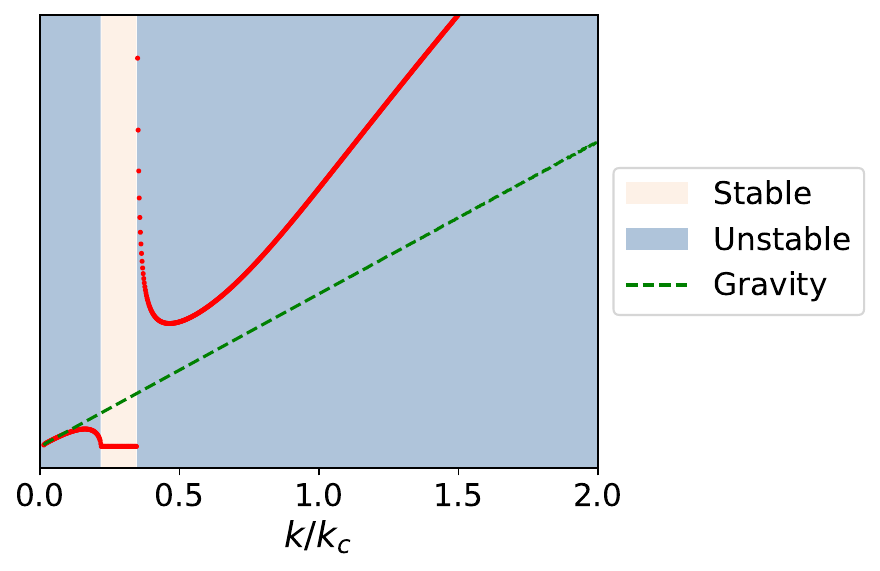}
    \caption{(Left) Dimensionless growth rate of instability in sector $\omega_+$: shows the enhanced growth rate of MI, relative to gravity waves. (Right)  The dimensionless growth rate of instability in $\omega_-$ sector. In both cases, we set $\delta=0.0001\, k_c$ and $|A_0|=\nu_o$. }
    \label{fig:growth_hydro}
\end{figure}

Moreover, in the unstable regions, the growth rate of MI also shows asymmetric behavior amongst the two sectors, as shown in Fig.~\ref{fig:growth_hydro}. We notice that for the $\omega_+$ sector the growth rate is enhanced relative to the gravity case, while for the $\omega_-$ sector the growth rate is suppressed relative to the gravity case. Also, for $\omega_-$, at intermediate and large $\bar k$, the growth rate is closer to the chiral limit.
Building on our analysis of modulation instability (MI) in odd gravity waves, a natural question arises: can light propagating in a nonreciprocal dielectric medium exhibit a dispersive parity-breaking phenomenon similar to odd viscosity? Furthermore, in the following sections, we seek to study how this parity-breaking effect interacts with nonlinear MI in such optical systems.

\section{Nonreciprocal 1D dielectric}\label{sec:optics}

Charges in a dielectric medium are bound in molecules and cannot move freely throughout the whole sample. Hence, their response to an external electric field causes a change in the charge distribution within each molecule, which induces an electric dipole moment in it. 



In particular, one-dimensional dielectric systems can be modeled by a group of damped harmonic oscillators that imitate molecules in a medium. The molecules have heavy nuclei that form a one-dimensional lattice along the $x$-axis, and each molecule has a mobile degree of freedom that enables it to move perpendicular to the lattice direction. When the spacing between adjacent atoms becomes extremely small, the model can be approximated by a continuous dielectric. In this case, the collection of molecular electric dipoles gives rise to a polarization field, which modifies the propagation of the electromagnetic field inside the medium.

When an electromagnetic wave propagates in this dielectric wire, the electromagnetic pulse moves forward, it interacts with the medium by initiating motion in the individual oscillators, which mimics the movement of electrons bound to nuclei in the dielectric. This links the evolution of the electromagnetic pulse in the medium to its effective polarization.



For the sake of simplicity, we will restrict the following analysis to linearly polarized electromagnetic waves. In this case, electric and magnetic fields are on the form $\vec E(\vec r,t)=E(x,t)\hat y$ and $\vec B(\vec r,t)=B(x,t)\hat z$
such that, they are connected through Faraday's law:
\begin{align}
    E_x+\frac{1}{c}B_t&=0\,. \la{eq:Faraday}
\end{align}
The electric field drives the oscillators such that their dynamics are described by
\begin{align}
   m_n \left(\ddot y_n+\gamma_n \dot y_n+\omega_{n}^2 y_n\right)=q_nE(x_n,t)\,, \la{eq:oscillators}
\end{align}
where $m_n$ is the oscillator effective mass, $\omega_{n}$ is the oscillator frequency, $q_n$ is the charge of the movable particle and $\gamma_n$ is the oscillator damping coefficient. Assuming that the oscillator are identical $\omega_n=\omega_0$ and separated by lattice-spacing $a$ along the $x$-direction, it is possible to express Eq.~(\ref{eq:oscillators}) in terms of the molecular electric dipole moment $p_n=qy_n$:
\begin{align}
   \ddot p_n +\gamma \dot p_n+\omega_0^2 p_n=\frac{q^2}{m}E(na,t)\,.
\end{align}
The polarization field is nothing but the electric dipole density. Thus, it can be defined by dividing the molecular electric dipole by the wire cross-section area and by the spacing between molecules and taking the continuum limit, that is,
\begin{align}
    P(x,t)=\lim_{a\rightarrow 0}\frac{q}{Aa}y_{n}(t)\,.
\end{align}
This limit is taken by setting $x=\lim_{a\rightarrow 0}na$ as the location of the infinitesimal electric dipole.

Therefore, for a continuum one-dimensional dielectric medium, the governing equations are given by
\begin{align}
    &P_{tt}+\gamma P_t+\omega_0^2 P=\frac{\omega_P^2}{4\pi} E\,, \la{eq:Polarization}
    \\
    &B_x+\frac{1}{c}E_t=-\frac{4\pi}{c}P_t\,, \la{eq:Ampere-Maxwell}
\end{align}
together with Faraday's law~(\ref{eq:Faraday}). The coefficient $\omega_P$ is often called plasmon frequency. The magnetic field can be eliminated by combining Eqs.~(\ref{eq:Faraday}, \ref{eq:Ampere-Maxwell}) which leads to 
\begin{align}
    c^2E_{xx}-E_{tt}=4\pi P_{tt}\,. \la{eq:ElectricField}
\end{align}

However, the presence of the damping coefficient $\gamma$ breaks the time-reversal symmetry, which leads to the absorption of the electromagnetic wave and a finite penetration depth in the wire. In this scenario, nonlinearities become irrelevant and the linear theory completely describes the system. Therefore, to study nonlinear media, one must ensure that $\gamma\rightarrow 0$, which preserves the time-reversal symmetry.

\subsection{Dispersive parity breaking in a dielectric} \label{sec:nonrec}

Following the intuition drawn from the hydrodynamic case, it should be possible to break the time-reversal symmetry keeping the system non-dissipative; as long as the combined PT symmetry is preserved. Eqs.~(\ref{eq:Polarization}) and (\ref{eq:ElectricField}) are symmetric under the parity transformation $x\rightarrow -x$. Therefore, terms such as $P_{tx}$, $P_{t,xxx}$, etc., violate parity and time-reversal symmetry, but preserve the combined PT symmetry~\footnote{In fact, any term with an odd order of time derivatives and an odd order of spatial derivatives will break T and P, but preserves PT.}. Modifying the dynamical equation for the polarization to accommodate for the simplest of these chiral terms gives us the following set of equations
\begin{align}
    &P_{tt}+\beta_{\text{odd}} P_{tx}+\omega_0^2 P=\frac{\omega_P^2}{4\pi} E\,, \la{eq:PolarizationChiral}
    \\
    &c^2E_{xx}-E_{tt}=4\pi P_{tt}\,.\nonumber
\end{align}
Note that equation~(\ref{eq:ElectricField}) since it is derived directly from Maxwell's equations. 

It is not hard to see that Eqs.~(\ref{eq:ElectricField}-\ref{eq:PolarizationChiral}) lead to wave propagation without any absorption, which satisfies the following dispersion relation:
\begin{equation}
    (c^2k^2-\omega^2)(\omega^2-\beta_{\text{odd}}\, \omega k-\omega_0^2)+\omega^2\omega_P^2 =0 \,.\label{Disper_odd}
\end{equation}
For $\beta_{\text{odd}}\neq 0$, the dispersion relation is not invariant under $\omega\rightarrow-\omega$ (time-reversal symmetry) or $k\rightarrow -k$ (parity symmetry) alone, nevertheless it remains the same if we combine both. Furthermore, note that parity and time-reversal become symmetries of the system if $\beta_{\text{odd}}$ change sign under them, that is, $\beta_{\text{odd}} \rightarrow  - \beta_{\text{odd}}$ for both P and T symmetries. 

\begin{figure}
    \centering
    \includegraphics[width=0.5\paperwidth]{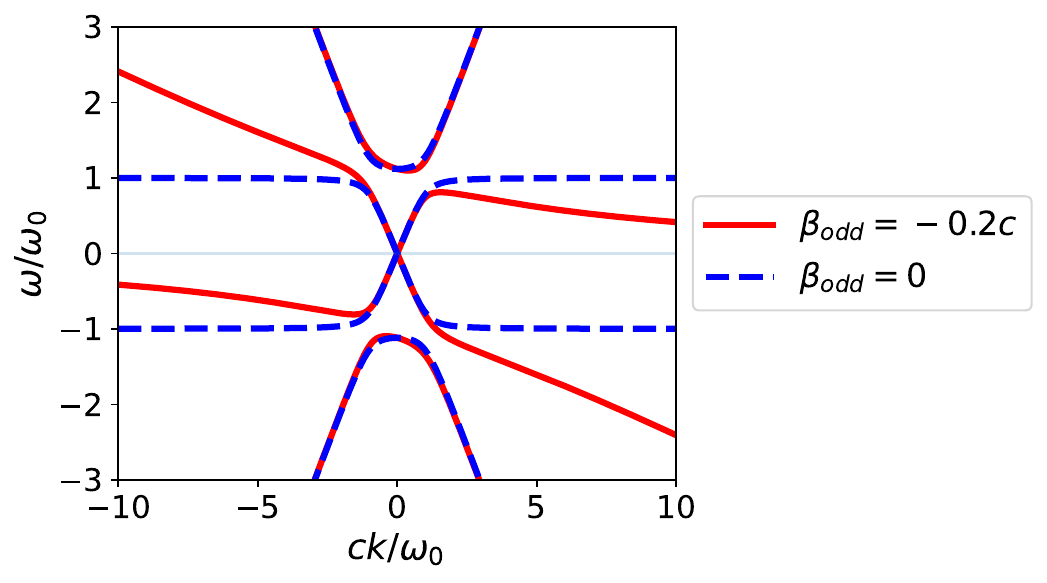}
    \caption{Dispersion relation for the parity broken dielectric media. The dashed curve is for the parity-symmetric dielectric. Scaling is such that $(\omega_P/\omega_0)=0.5$. }
    \label{fig:disper}
\end{figure}

In this medium, the chiral nature is evident, since the presence of the coefficient $\beta_{\text{odd}}$ skews the dispersion relation and creates an asymmetry between left and right moving modes, as seen in Fig.~\ref{fig:disper}. This effect is more pronounced in the two middle bands at large $k$. In this limit, polarization and electromagnetic waves decouple from each other and the middle bands describe the polarization waves, which asymptotically satisfy the the dispersion relation $\omega^2-\beta_{\text{odd}}\, \omega k-\omega_0^2=0$, or 
\begin{align}
    \omega\sim \tfrac{1}{2}\left(\beta_{\text{odd}} k\pm\sqrt{\beta_{\text{odd}}^2k^2+4\omega_c^2}\right).
\end{align}

Equivalently, the top and the bottom bands describe the uncoupled electromagnetic waves at large $k$, which satisfy the dispersion relation $\omega=\pm c|k|$. For these bands, the parity breaking shows up for small $k$, when the coupling between polarization and electromagnetic waves is substantial.   

\subsection{Heuristics of the Chiral Term}

The term $\beta_{\text{odd}} P_{tx}$ was phenomenologically introduced in the setup, however, it can also be obtained from a microscopic model. In the oscillator picture, we find that
\begin{align}
    \beta_{\text{odd}} P_{tx}\approx \frac{\beta_{\text{odd}}\, q}{Aa}\left(\frac{\dot y_{n+1}-\dot y_{n-1}}{2a}\right).
\end{align}
For $\beta q>0$, this corresponds to a force acting on the oscillator $n$, which is along the direction of motion of its right neighbor and opposite to the motion of its left neighbor. This automatically breaks parity, since each molecule feels a drag force coming from the adjacent right neighbor and a push to oppose the motion of the adjacent left neighbor. 

Even though these forces could be derived directly from a Lagrangian model, obtained from the term 
\begin{align}
    \mathcal L_{PT}=\frac{m\beta_{\text{odd}}}{2a}y_n\dot y_{n+1}\,,
\end{align}
it is not clear to us what underlying mechanism allows for such a term. Identifying this can lead to interesting non-reciprocal effects in 1D polariton systems since this model can be viewed as a classical version of the polariton systems.

\subsection{Kerr Nonlinearity}\label{sec:kerr}

Nonlinearities can be introduced into the model by adding anharmonic terms to the restoring force. The simplest term that preserves the oscillator's symmetry is a cubic restoring force: $F_{Kerr}=-\kappa y_n^3$. This is generally referred to as the Kerr medium and allows for the generation of higher harmonics. The polarization dynamics now becomes  
 \begin{equation}
      P_{tt}+\beta_{\text{odd}} P_{tx}+\omega_0^2 P+\alpha P^3=\frac{\omega_P^2}{4\pi} E\,. \label{eq:F_EOM_NL}
\end{equation}
In the optics case, the Maxwell equation~(\ref{eq:ElectricField}) remains linear, while nonlinearities emerge in the polarization dynamics. By applying the method of multiple scales, as previously used in the hydrodynamic case, we can analyze modulation instability (MI) in this classical polariton system.

After introducing the same variables defined in Eq.~(\ref{multiscale}), and applying the method of multiple scales to Eqs.~(\ref{eq:ElectricField}) and  (\ref{eq:F_EOM_NL}), we obtain the NLSE corresponding to the polariton wave envelope to be
\begin{equation}
    i\partial_{\tau} P_{11} +\tfrac{1}{2}\, \omega''(k)\partial_{\xi}^{2}P_{11} -M |P_{11}|^{2}P_{11}=0 \label{NLS}
\end{equation}
where 
\begin{align}
    M=\frac{3 \,\alpha\,\omega_P^2\, \omega^3}{\omega_P^2\, \omega^2 (\omega^2 + \omega_0^2) +(\omega^2 +c^2 k^2)(\omega_0^2+\beta_{\text{odd}}\,\omega k -\omega^2)^2}\,.
\end{align}
For calculation details, we refer to Appendix~\ref{app:optics}. Since $\alpha>0$, the sign of $M$ is determined by the sign of $\omega$. This means that the condition $LM \geq 0$ can be recast as
\begin{equation}
    \omega\frac{d^2\omega}{dk^2}\geq0.
    \label{eq:opticsstability}
\end{equation}

The stability criterion in this case is simpler than in the hydrodynamic case, as it only depends on the nature of the dispersion relation. When $\beta_{\text{odd}} = 0$, the top and bottom bands are stable under modulation instability (MI) as they satisfy Eq.(~\ref{eq:opticsstability}). The two central bands, on the contrary, are unstable for all $ k \neq 0$ modes in this parity-preserving case. When $\beta_{\text{odd}} \neq 0$, while the top and bottom bands remain stable, new stability pockets appear in the middle bands.

\begin{figure}[H]
       \centering \includegraphics[width=0.35\paperwidth]{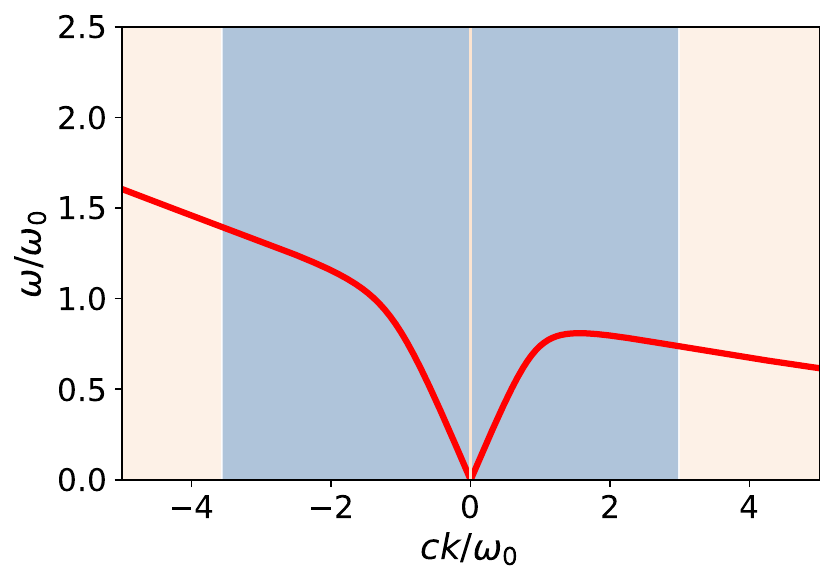}
\includegraphics[width=0.35\paperwidth]{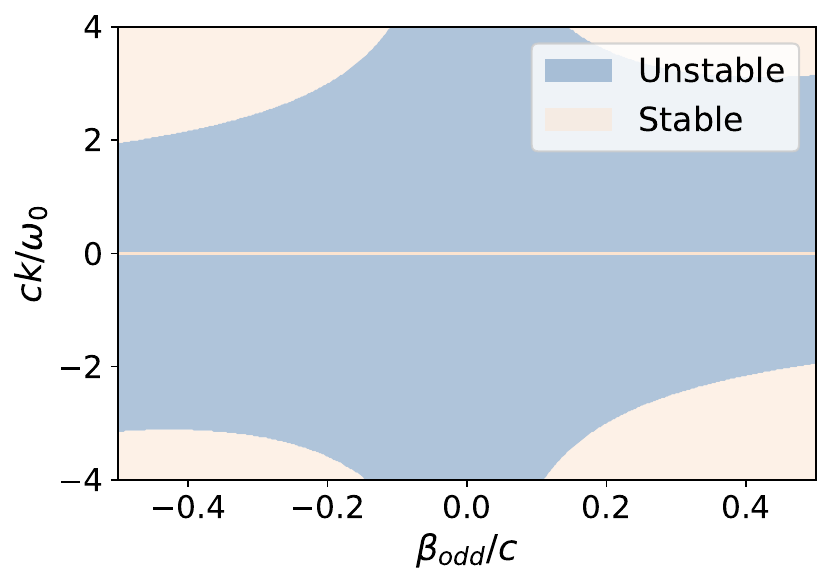}
    \caption{(Left) Lower Polariton band for parity broken case, with $\beta_{\text{odd}} =-0.2c$. For the two sectors, the band curvature changes sign in the regions $c k/\omega_0 < -3.579$ and $c k/\omega_0>2.989$, which leads to the stabilization of modes in the respective sectors.
   (Right) Stability in the $\beta_{\text{odd}}$-k space of the lower polariton mode, in $\delta \rightarrow 0$ limit. Dark regions are unstable and light regions are stable. Stability is calculated in vanishing bandwidth limit ($\delta=0$). The parameters of the model considered are the same as Fig.~\ref{fig:disper}.
    }
    \label{fig:instab_odd}
\end{figure}

\begin{figure}[H]
        \centering\includegraphics[width=0.5\paperwidth]{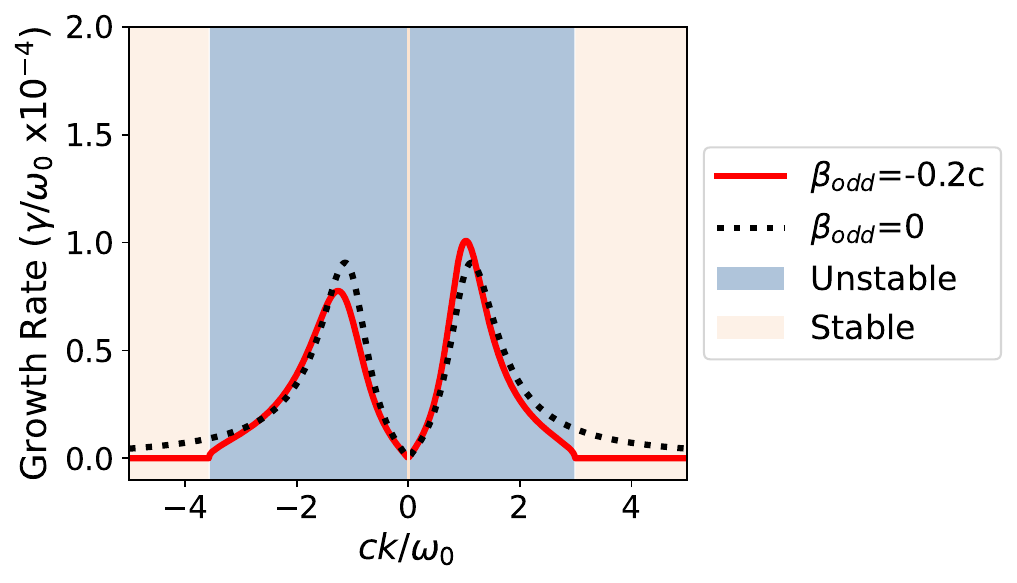}
    \caption{Dimensionless growth rate for MI in parity broken dielectric, with $\beta_{\text{odd}}=-0.2c$. Parameters considered are the same as Fig.~\ref{fig:disper}, with $\delta=0.0001\,\omega_0/c$ and $|A_0|=\omega_0/\sqrt{\alpha}\,$. 
     }
\label{fig:odd-growth}
\end{figure}

In particular, we focus on the lower polariton band (the positive-frequency middle band) in the parity-broken case with $\beta_{\text{odd}} = -0.2c$ (see Fig.~\ref{fig:instab_odd} left panel). For this case, the dispersion relation satisfies $\omega > 0$ while $\omega''(k)$ is positive for $k < -3.579 \,\omega_0/c$ and $k > 2.989 \,\omega_0/c$, resulting in the stabilization of modes in these regions. In the region where modes are unstable, parity breaking modifies the growth rate of instability in an asymmetric way (differently for $k > 0$ and $k < 0$) as is evident in Fig.~\ref{fig:odd-growth}. Furthermore, in the ($\beta_{\text{odd}}$ vs $k$) plot, shown in Fig.~\ref{fig:instab_odd} right panel, the regions of stability appear only when $\beta_{\text{odd}} \neq 0$ at certain critical values of $k$, which are determined by solving $\omega''(k) = 0$. Note that $k = 0$ is trivially stable since the nonlinear term vanishes for this case.

\section{Discussion}
In this work, we studied the interplay of parity breaking symmetry and the modulation instability (MI) in two distinct physical systems, fluid dynamics and optics, with characteristic wavelengths differing in principle by orders of magnitude. For the hydrodynamics system, we considered the nonlinear odd gravity waves: a 2D free surface problem with gravity $g$ and odd viscosity $\nu_o$. Using the method of multiple scales, we derived the nonlinear Schrödinger equation (NLSE) for the slowly evolving envelope dynamics. We performed a linear stability analysis for this NLSE and showed that the stability criteria for the left-moving and right-moving NLSE are very different due to the parity breaking in the system. In the intermediate regime, where gravity's effects are comparable to parity breaking, gravity and odd viscosity-induced capillary waves interact resonantly, resulting in a narrow window of stability around this resonance. At resonance, our counting scheme breaks down, and a different counting scheme is required. This situation is reminiscent of the resonant interaction between gravity waves and surface tension-induced capillary waves. However, in the case of gravity and surface tension, both left and right propagating waves resonate, unlike the present case where the resonance occurs in only one sector. {\it One interesting consequence of such chirality-dependent stability is that MI-triggered solitons propagate in only one direction in these nonlinear systems.}

Following the intuition drawn from the hydrodynamical case, we construct a model for a non-reciprocal dielectric in 1D that realizes analogous parity-broken effects. The term that breaks parity couples the velocity of one oscillator to the position of the next oscillator. In the continuum limit, this term includes a spatial and a time derivative, which is odd under parity (P) and time reversal (T) but invariant under the combined PT action. The continuum equations for electromagnetic waves passing through this medium contain a dispersive parity-breaking term similar to the odd viscosity term found in fluids.

We emphasize that, unlike usual non-reciprocal media, which are not Hamiltonian, our model system is energy-conserving and Hamiltonian.  The presence of the second-order temporal derivatives of the coupled equations for the polarization and electric field results in four branches for the dispersion relation. With a cubic nonlinearity in the polarization equation combined with the parity-breaking effects, we derive the nonlinear Schrödinger equation (NLSE) using the method of multiple scales and analyze the stability criterion as a function of the parameters and the wavenumber. In contrast to the NLSE for odd gravity waves, the stability criterion in the optics case depends only on the form of the dispersion relation. The non-reciprocity leads to stability for large wavenumbers, where the effects of parity breaking are most dominant (see Fig.~\ref{fig:instab_odd}).   

In conclusion, we believe the phenomena discussed in the work are likely observable in other fields of physics with the interplay of parity breaking and MI being natural, such as astrophysics, plasma physics, non-reciprocal metamaterials, and Bose-Einstein condensates of dipolar gases. As such, our work could be the start of a direction in the investigation of chirality-dependent instabilities and their consequences in a wide class of systems. 

\section{Acknowledgements}
We thank Vinod Menon for motivating this work. This work is supported by NSF CAREER Grant No. DMR-1944967 (SG and SS). SS and SG were also supported in part by NSF OMA-1936351.

\section{Appendix}
\subsection{Details for envelope solutions in parity-odd fluid } \label{app:hydro}
The deep-water surface system Eq.(\ref{eq:bulkchi})-Eq.(\ref{eq:DBC}), can be expressed in terms of envelope variables, defined in Eq.(\ref{multiscale}), that is,
\begin{equation*}
    \sigma=kx-\omega t\,,\qquad \xi= \epsilon (x-c_g t )\,,\qquad \tau =\epsilon^2 t\,.
\end{equation*} 
In these variables, the bulk equation $(-\infty\leq \chi\leq0)$ becomes
\begin{align}
      & k^2\theta_{\sigma\sigma}+(1+k^2\eta_\sigma^2)\theta_{\chi\chi}-2k^2\p_\sigma(\eta_\sigma\theta_\chi)+k^2\eta_{\sigma\sigma}\theta_{\chi}+2k\epsilon\big[\theta_{\sigma\xi}+\eta_\sigma\eta_\xi\theta_{\chi\chi}\nonumber
      \\
      &-\p_\xi(\eta_\sigma\theta_\chi)-\eta_\xi\theta_{\sigma\chi}\big]+\epsilon^2\left[\theta_{\xi\xi}+\eta_\xi^2\theta_{\chi\chi}-2\p_\xi(\eta_\xi\theta_\chi)+\eta_{\xi\xi}\theta_{\chi}\right]=0,
       \end{align}
whereas the kinematic boundary condition can be written as 
\begin{align}
    &k^2\eta_\sigma\theta_\sigma-\omega\eta_\sigma-(1+k^2\eta_\sigma^2)\theta_\chi+\epsilon\big[k(\eta_\sigma\theta_\xi+\eta_\xi\theta_\sigma)\nonumber
    \\   &-c_g\eta_\xi-2k\eta_\xi\eta_\sigma\theta_\chi\big]+\epsilon^2\left[\eta_\tau+\eta_\xi\theta_\xi-\eta_\xi^2\theta_\chi\right]=0\,,  \qquad  \qquad \chi=0 .
    \end{align}
On the other hand, the no-penetration condition at the infinity does not get modified, and it remains 
\begin{align}
      &\theta_\chi=0\,, \qquad \qquad \qquad \qquad     \chi=-\infty .\label{eq:bottomBC}
\end{align}

In addition to these, we have the Bernoulli equation taken at the surface, where the effects of odd viscosity are encoded as the modified dynamical pressure term. This is in contrast to the ideal fluids case where the pressure is simply zero. In terms of the envelope variables, this becomes
\begin{align}
    &-\omega(\theta_\sigma-\eta_\sigma\theta_\chi)+\tfrac{1}{2}\big[k^2\theta_\sigma^2+(1+k^2\eta_\sigma^2)\theta_\chi^2-2k^2\eta_\sigma\theta_\sigma\theta_\chi\big]+g\eta+\epsilon\big[c_g(\eta_\xi\theta_\chi\-\theta_\xi)+k\theta_\sigma\theta_\xi\nonumber
    \\ 
    &-k\theta_\chi\left(\eta_\sigma\theta_\xi+\eta_\xi\theta_\sigma-\eta_\sigma\eta_\xi\theta_\chi\right)\big]+\epsilon^2\left[\theta_\tau-\eta_\tau\theta_\chi+\tfrac{1}{2}\left(\theta^2_\xi+\eta-\xi^2\theta_\chi^2-2\eta_\xi\theta_\xi\theta_\chi\right)\right]=\nonumber
    \\
    &-\frac{2\nu_o}{\sqrt{1+(k\eta_\sigma+\epsilon\eta_\xi)^2}}(k\partial_\sigma+\epsilon\partial_\xi)\left[\frac{\omega\eta_\sigma+c_g\eta_\xi-\epsilon^2\eta_\tau}{\sqrt{1+(k\eta_\sigma+\epsilon\eta_\xi)^2}}\right],\qquad\qquad \quad \chi=0\,.
\end{align}
Even though the odd viscosity term in the last line looks horrendous, it simplifies quite substantially when we consider that $\eta,\theta\sim \mathcal O(\epsilon)$.
We seek a perturbative expansion for this system of equations of the type:
\begin{equation}
 \eta=\sum_{j=1}^\infty \,\epsilon^j\sum_{n=-j}^j\eta_{jn} q^n\,, \qquad \qquad \theta=\sum_{j=1}^\infty \,\epsilon^j\sum_{n=-j}^j\theta_{jn}q^n\,,   
\end{equation}
which will lead to the envelope dynamics for the slow degrees of freedom, as prescribed by Eq.(\ref{eq:multiscale2}). The equations at order $\mathcal{O}(\epsilon q)$ can be written as
\begin{align}
	\p_{\chi}^2\theta_{11}&=k^2 \theta_{11},\quad -\infty\leq \chi\leq 0\\
	 \p_\chi \theta_{11}&=0,              \qquad      \quad    \quad \chi=-\infty\\
	 -i\omega \eta_{11}&= \p_\chi \theta_{11}\,\,\,\qquad \chi=0\\
	  -i\omega \theta_{11}+g \eta_{11} &= 2\nu_o \omega k \eta_{11}, \quad \chi=0\\
	   \theta_{10}&=A_0(\xi,\tau) \quad  -\infty\leq \chi\leq 0\\
	   \eta_{10}&=0  
\end{align}
From here on, we will consider $k>0$. In fact, the analysis for $k<0$, can be recovered under the replacement $K\rightarrow -k$ and $\omega\rightarrow-\omega$. The first two equations can be satisfied by the solution
\begin{align}
	&\theta_{11}=A(\xi,\tau)e^{k \chi},\qquad \eta_{11}=\frac{i \omega}{g-2\nu_o\omega k}A(\xi,\tau)=i (k/\omega) A(\xi,\tau)\, 
\end{align}where $A$ is the slowly-varying unknown carrier wave-train amplitude. The dispersion of the general system is also obtained in this order and gives us
\begin{align}
    \omega^2 +2\nu_o k^2 \omega -gk=0\,.
\end{align}
Since we have fixed $k>0$, this leads to the two sectors $\omega_{\pm}$ (right and left movers), be defined by the two branches of the above dispersion.
\begin{align}
    \omega_{\pm}=-\nu_ok^2\pm\sqrt{gk+\nu_o^2k^4}
\end{align}
In the following, we use the general $\omega$ which has two possible solutions $\omega_{\pm}$ for $k>0$. In the final NLSE, we can analyze the dynamics for both solutions. We solve for $\mathcal{O}(\epsilon^2 )$ equations at the free surface. The higher order bulk involves more unknown functions.  Comparing the harmonics, we obtain the following equations for the bulk solution:
\begin{align}
&\mathcal{O}(\epsilon^2 q^0):\quad    \p_{\chi}^2\theta_{20}=0\,,\\
&\mathcal{O}(\epsilon^2 q^1):\quad	\p_{\chi}^2\theta_{21}-k^2 \theta_{21}+2ik\p_\xi A e^{k \chi}=0\,,\\
&\mathcal{O}(\epsilon^2 q^2):\quad	\p_{\chi}^2\theta_{22}-4k^2 \theta_{22}+\frac{3ik^4 A^2 e^{k\chi}}{\omega}=0\,.	
\end{align}
The associated solutions for the bulk subject to the boundary condition 	$\p_\chi\theta_{jn}\Big|_{\chi=-\infty}=0$ at this order is given by
\begin{align}
	\theta_{21}&=\left[A_2(\xi,\tau)e^{k\chi}-i (\p_\xi A) \chi e^{k \chi}\right],
 \\
	 \theta_{22}&= A_3(\xi,\tau)e^{2k \chi}+ i \left(\frac{k^2 A^2}{\omega}\right) e^{ k \chi}\,,
  \\
	 \theta_{20}&=A_4(\xi,\tau)\,,
\end{align}
where $A_2, A_3, A_4$ are unknown functions of $\xi$ and $\tau$. Using the boundary conditions we can express them in terms of the linear order envelope $A(\xi,\tau)$.

The zeroth harmonic trivially satisfies the kinematic boundary at this order. Comparing the first harmonic and second harmonic at this order, we obtain 
\begin{align}
&\mathcal{O}(\epsilon^2 q^1):	-i\omega \eta_{21}=i\left(\frac{c_g k-\omega}{\omega}\right)\p_\xi A(\xi, \tau)+k A_2\,,
\\
&\mathcal{O}(\epsilon^2 q^2):	-2i\omega \eta_{22}=\left(\frac{2ik^3 }{\omega}\right)A^2+2k A_3(\xi,\tau)\,.
\end{align}
From the dynamical boundary condition, we obtain
\begin{align}
	&\mathcal{O}(\epsilon^2 q^0):\quad c_g \p_\xi A_0=g\,\eta_{20}\,,
 \\
	&\mathcal{O}(\epsilon^2 q^1):\quad (g-2\nu_o\omega k)\eta_{21}=i \omega A_2+\left(\frac{2c_g \nu_o k^2}{\omega}+2\nu_o k+c_g\right)\p_{\xi}A\,,
 \\
	&\mathcal{O}(\epsilon^2 q^2):\quad (g-8\nu_o\omega k)\eta_{22}=2i\omega A_3-k^2A^2\,.
 \end{align}
The equation involving $\eta_{21}$ and $A_2$ are not independent and consequently, we only obtain the following combination as a condition
\begin{align}
	\omega \,\eta_{21}- i k A_2=\left(1-\frac{c_g k}{\omega}\right)\p_\xi A\,. \label{A2_eta21}
 \end{align}

The higher-order equations depend only on the aforementioned combination or its derivatives. Solving the equations appearing in the second harmonic, we obtain,
\begin{align}
	\eta_{22}=\frac{- k^3}{\omega(\omega+6\nu_ok^2)}A^2,\qquad A_3=\frac{-6i\nu_o  k^4}{\omega(\omega+6\nu_ok^2)}A^2.
 \end{align}
For one of the branches, we observe the second-harmonic resonance in the solution $\omega_-(k)+6\nu_ok^2=0$ due to interactions between the gravity waves and odd surface waves. At the resonance value of $k$, the present counting scheme breaks down and one needs to choose a different counting scheme. The remainder of this analysis assumes that we are away from this resonance point.  At $\mathcal{O}(\epsilon^3 q^0)$ we get for the bulk
  \begin{align}
      \mathcal{O}(\epsilon^3 q^0):\p_{\chi\chi}^{2}\theta_{30} +\p_{\xi\xi}^2\theta_{10} + \left( 2ik\eta_{1-1}\p_{\chi\xi}^2\theta_{11} +c.c.\right) -k^2\left( \eta_{2-1}\p_{\chi}\theta_{11} + \eta_{11}\p_{\chi}\theta_{2-1} + c.c. \right)=0,\nonumber
  \end{align}
  where $c.c.$ refers to complex conjugation.
Solving the inhomogeneous bulk equation, we get
  \begin{align}
      \theta_{30}=\left(\frac{-\p_{\xi\xi}^2A_0}{2}\right)\chi^2 + \left(\frac{k^2c_g}{\omega^2} + \frac{k^2 \chi}{\omega}\right) e^{k\chi} \p_{\xi\xi}^2 |A|^2\,.
  \end{align}
 However, the boundary condition imposes ($     \p_{\chi}\theta_{30}=0 $ as $\chi= -\infty$) leading to the constraint $A_0=0$. We further need the bulk equation at $\mathcal{O}(\epsilon^3 q^1)$, which will be used to eliminate $\theta_{31}$ fields in the surface equations
	\begin{align}
		\mathcal{O}(\epsilon^3 q^1):	\p_{\chi\chi}^2\theta_{31}-k^2\theta_{31}=\frac{-6ik^4}{\omega}A^* A_3 e^{2k \chi}-2i k\p_\xi A_2 e^{k \chi}-\p_{\xi\xi}^2A(2k \chi +1)e^{k\chi}\,.
			\end{align}
Solving this bulk equation gives us another unknown function $A_5$ and the full solution is given by
			\begin{align}
				\theta_{31}=A_5 e^{k \chi}-\frac{12 \nu_ok^6}{\omega^2(\omega+6\nu_ok^2)}A|A|^2 e^{2k \chi}-i \p_\xi A_2 \chi e^{k \chi}-\chi^2 e^{k\chi} \p_{\xi\xi}^2A/2.
			\end{align}
The last set of equations we need to consider is the two boundary conditions at $\mathcal{O}(\epsilon^3 q^1)$. After simplifying these equations using the lower order solutions, we obtain
	\begin{align}
	&- \frac{\omega^2}{k}\eta_{31}+i \omega A_5+\p_\tau A+ i \frac{c_g \omega}{k} \p_\xi \eta_{21}+\frac{\omega}{k} \p_\xi A_{2}-\left(\frac{i k^4 }{\omega }-\frac{4 i k^4 }{6 k^2 \nu_o+\omega }\right)A |A|^2=0\,,
 \\
	&\frac{\omega^2}{k}\eta_{31}-i\omega A_5+\left(1+\frac{2\nu_ok^2}{\omega}\right)\p_\tau A+2i\nu_o(\omega+kc_g) \p_\xi \eta_{21} -c_g\p_{\xi}A_2+\frac{2i\nu_o k c_g}{\omega}\p_{\xi\xi}^2A\nonumber
 \\
	&+i k^4 \left(\frac{4 \nu_o k^2}{\omega ^2}+\frac{12 \nu_o k^2}{\omega(\omega+6 k^2 \nu_o) }+\frac{1}{\omega }\right)A|A|^2=0\,.
	\end{align}

We can add the equations above to simultaneously eliminate $\eta_{3,1}$ and $A_5$. Further using the linear combination Eq.(\ref{A2_eta21}), and the dispersion, we find a consistency condition in the form of the non-linear Schr\"odinger equation for the envelope as discussed in the main text,
	\begin{align}
	i\p_\tau A+L\p_{\xi\xi}^2A- M A|A|^2=0. \la{eq:NLS}
\end{align}
where 
\begin{align}
    L=\frac{1}{2}\frac{d^2\omega}{dk^2},\qquad M=\frac{2k^4}{\omega}\left( 1-\frac{3\nu_ok^2\omega}{(\omega+\nu_o k^2)(\omega+6 \nu_o k^2)}\right).
\end{align}
This is the final result that is presented in the main text.

\subsection{Envelope dynamics in the parity-odd dielectric medium} \label{app:optics}

In this section, we present the calculation details for the envelope dynamics in the 1D light-matter dielectric system with broken parity. The starting point is the set of following equations of motion:
\begin{align*}
    &P_{tt}+\beta_{\text{odd}} P_{tx}+\omega_0^2 P+\alpha P^3=\frac{\omega_P^2}{4\pi} E\,
    \\
    &c^2E_{xx}-E_{tt}=4\pi P_{tt}\,.
\end{align*}
In terms of the multiple scales variables introduce in Eq.~(\ref{multiscale}), that is,
\begin{equation*}
    \sigma=kx-\omega t\,,\quad \xi= \epsilon (x-c_g t )\,,\quad \tau =\epsilon^2 t\,,
\end{equation*}
these dynamical equations become
 \begin{align}
    &(\omega^2 -c^2 k^2)E_{\sigma\sigma} +4\pi \omega^2  P_{\sigma\sigma} + \epsilon\left[ 2(c_g\,\omega -c^2 k) E_{\sigma\xi} + 8\pi c_g \omega P_{\sigma\xi}\right]+\epsilon^{2} [-2\omega E_{\sigma\tau}\nonumber \\
    &    +( c_g^{2} -c^2)E_{\xi \xi}-8\pi\omega P_{\sigma \tau} + 4\pi c_g^{2}P_{\xi\xi} ] +O(\epsilon^3)=0\,,
    \\
    &(\omega^2 -\beta_{\text{odd}}\, \omega k)P_{\sigma\sigma} +\omega_0^{2} P +\alpha P^3 -\frac{\omega_P^2}{4\pi} E  +\epsilon\, [ 2\omega c_g -\beta_{\text{odd}} (\omega +k c_g)] P_{\sigma \xi} \nonumber 
    \\
    &+\epsilon^{2}[(\beta_{\text{odd}} k-2\omega)P_{\sigma \tau}  +c_g (c_g - \beta_{\text{odd}}) P_{\xi\xi}] +O(\epsilon^3)=0\,.
\end{align}

Using the intuition from the hydrodynamic case, we only need terms up to $\mathcal O(\epsilon^2)$ in this expansion, since this will produce all the terms at $\mathcal O(\epsilon^3)$ when we impose that $P,E\sim\mathcal O(\epsilon)$. Expanding $E$ and $P$ perturbatively as
\begin{align*}
    E=\sum_{j=1}^{\infty} \epsilon^j E_j\,,\qquad     P=\sum_{j=1}^{\infty} \epsilon^j P_j,
\end{align*}
and arranging the system order by order in $\epsilon$ we get
\begin{align}
\mathcal{O}(\epsilon):\qquad  &(\omega^2-c^2k^2)E_{1\sigma\sigma}+4\pi\omega^2P_{1\sigma\sigma}=0, \la{eq:E1}
\\
&(\omega^2-\beta_{\text{odd}}\omega k)P_{1\sigma\sigma}+\omega_0^2P_{1}-\frac{\omega_p^2}{4\pi}E_1=0\,, \la{eq:P1}
\\
\nonumber
\\
\mathcal{O}(\epsilon^2): \qquad
&(\omega^2-c^2k^2)E_{2\sigma\sigma}+4\pi\omega^2P_{2\sigma\sigma}+2(\omega c_g-c^2k)E_{1\sigma\xi}+8\pi\omega c_g P_{1\sigma\xi}=0,
\\
&(\omega^2-\beta_{\text{odd}}\,\omega k)P_{2\sigma\sigma}+\omega_0^2P_{2}-\frac{\omega_p^2}{4\pi}E_2+(2\omega c_g-\beta_{\text{odd}}(k c_g+\omega))P_{1\sigma\xi}=0\,,
\\
\nonumber
\\
\mathcal{O}(\epsilon^3):\qquad
&(\omega^2-c^2k^2)E_{3\sigma\sigma}+4\pi\omega^2P_{3\sigma\sigma}+2(\omega c_g-c^2k)E_{2\sigma\xi}+8\pi\omega c_g P_{2\sigma\xi}\nonumber
\\
&-2\omega E_{1\sigma\tau}+(c_g^2-c^2)E_{1\xi\xi}-8\pi\omega P_{1\sigma\tau}+4\pi c_g^2P_{1\xi\xi}=0,\la{third_1}
\\
&(\omega^2-\beta_{\text{odd}}\,\omega k)P_{3\sigma\sigma} -\frac{\omega_p^2}{4\pi}E_3+(2\omega c_g-\beta_{\text{odd}}\,(k c_g+\omega))P_{2\sigma\xi}\nonumber
\\
&+(\beta_{\text{odd}}\,k-2\omega)P_{1\sigma\tau}+c_g(c_g-\beta_{\text{odd}})P_{1\xi\xi}+\omega_0^2P_{3}+\alpha P_1^3=0.\la{third_2}
\end{align}

Let us we further expanding each $\epsilon$-order into harmonic modes as
\begin{align*}
    E_j=\sum_{n=-j}^{j}  E_{jn}(\xi,\tau)e^{in\sigma}\,,\qquad       P_j=\sum_{n=-j}^{j}  P_{jn}(\xi,\tau)e^{in\sigma}.
\end{align*}
Due to reality condition of $E$ and $P$, we have $E_{j,-n}=E_{j,n}^*$ and $P_{j,-n}=P_{j,n}^*$. Note that under this scheme, higher harmonics can only be generated perturbatively. Therefore, the leading order equations, that is, Eqs.~(\ref{eq:E1}-\ref{eq:P1}) can be written as
\begin{align}
    &\sum_{n=-1}^{1}\left[ n^2(k^2c^2-\omega^2)E_{1n}-4\pi n^2\omega^2P_{1n}\right]e^{in\sigma}=0\,,
    \\ 
    &\sum_{n=-1}^{1}\left[ (n^2\omega(\beta_{\text{odd}}k-\omega)+\omega_0^2)P_{1n} -\frac{\omega_p^2}{4\pi}E_{1n}\right]e^{in\sigma}=0\,.
\end{align}
The zeroth harmonic mode gives us $\omega_0^2 P_{10}=\frac{\omega_p^2}{4\pi}E_{10}$. However, this corresponds to a uniform and constant electric (polarization) field in the medium. Since we are interested in the field propagation, we can set $E_{10}=P_{10}=0$. 

The rest of the leading order equations give us the dispersion relation inside the medium and how the electric field relates to the polarization field:
\begin{align}
&(\omega^2-c^2k^2)(\omega_0^2+\beta_{\text{odd}}\,\omega k-\omega^2)+\omega^2\omega_p^2=0\,, \nonumber
\\
&E_{11}=\frac{4\pi}{\omega_p^2}(\omega_0^2+\beta_{\text{odd}}\,\omega k-\omega^2)P_{11}=\frac{4\pi\omega^2}{c^2k^2-\omega^2}P_{11}\,.
\end{align}

Plugging the solutions of $\mathcal O(\epsilon)$ into the equation for order $\mathcal O(\epsilon^2)$, we find that the equations do not depend on $E_{11}$, $P_{11}$. Since nonlinear terms only appear in the third-order equations, the terms $P_2$ and $E_2$ do not contribute for the NLSE and can be set to zero without any loss of generality. From the same argument, we can also set $E_{32}=E_{30}=P_{32}=P_{30}=0$, leaving us with the following equations to consider
\begin{align}
  &(c^2k^2-\omega^2)E_{33}-4\pi\omega^2 P_{33}=0\,, \la{order3_harm3_first}
  \\
  &9(\omega\beta_{\text{odd}}\,k-\omega^2)P_{33}+\omega_0^2P_{33}-\frac{\omega_p^2}{4\pi}E_{33}+\alpha P_{11}^3=0\,,\la{order3_harm3_second}  
\end{align}
which determines the generation of the third harmonics in the medium in terms of the incident wave envelope, and
\begin{align}
    &(c^2k^2-\omega^2)E_{31}-4\pi\omega^2P_{31}-2i\omega E_{11\tau}+(c_g^2-c^2)E_{11\xi\xi}-i8\pi\omega P_{11\tau}+4\pi c_g^2 P_{11\xi\xi}=0\,, \la{order3_harm1_first}
    \\
    &\frac{\omega^2\omega_P^2}{c^2k^2-\omega^2}P_{31}-\frac{\omega_p^2}{4\pi}E_{31}+3\alpha P_{11}^2 P_{1,-1} +i(\beta_{\text{odd}}k-2\omega)P_{11\tau}+c_g(c_g-\beta_{\text{odd}})P_{11\xi\xi}=0\,. \la{order3_harm1_second}
\end{align}

After some algebra, the set of Eqs.~(\ref{order3_harm1_first}-\ref{order3_harm1_second}) can be expressed in form of the NLSE:

\begin{align}
    \left(i\partial_{\tau}+\frac{1}{2}\frac{d^2\omega}{dk^2}\partial_{\xi}^2\right)P_{11} -\frac{3\alpha\omega^3\omega_p^2}{(\omega^2\omega_p^2)(\omega^2+\omega_0^2)+(\omega_0^2+\beta_{\text{odd}}\omega k-\omega^2)^2(\omega^2 + c^2k^2)} |P_{11}|^2 P_{11}=0\,.
\end{align}
Since $\alpha >0$, the envelope solution is stable to MI, in zero bandwidth limit, when: 
\begin{align*}
    \omega \frac{d^2\omega}{dk^2} \geq 0\,.
\end{align*}

\bibliographystyle{unsrt} 
\bibliographystyle{SciPost_bibstyle} 
\bibliographystyle{my-refs.bst}
\bibliography{oddviscosity-bibliography.bib}
\nolinenumbers

\end{document}